\documentclass[10pt,journal,compsoc]{IEEEtran}

\usepackage{ifpdf}
\ifpdf
\else
\fi

\ifCLASSOPTIONcompsoc
  \usepackage[nocompress]{cite}
\else
  \usepackage{cite}
\fi
\ifCLASSINFOpdf
  \usepackage[pdftex]{graphicx}
  \graphicspath{{figures/}}
  \DeclareGraphicsExtensions{.pdf,.jpeg,.png}
\else
  \usepackage[dvips]{graphicx}
  \graphicspath{{../eps/}}
  \DeclareGraphicsExtensions{.eps}
\fi
\usepackage{amsmath}
\usepackage{mathtools}
\usepackage{amsfonts}
\usepackage{color}
\usepackage{algorithm}
\usepackage{algpseudocode}
\usepackage{booktabs}
\usepackage[table,xcdraw]{xcolor}
\usepackage{makecell}
\usepackage{array}

\usepackage{mdwmath}
\usepackage{mdwtab}

\usepackage{eqparbox}
\begin{document}
%
\title{MPMNet: A Data-driven MPM Framework for Dynamic Fluid-solid Interaction}
%
%
%
%

\author{Jin Li,
         Yang Gao\thanks{ J. Li and Y. Gao contributed equally and should be regarded as co-first authors.},
         Ju Dai,
         Shuai Li,
        Aimin Hao,
        Hong Qin,~\IEEEmembership{Member,~IEEE}
\IEEEcompsocitemizethanks{

\IEEEcompsocthanksitem J. Li, Y. Gao, S. Li and A. Hao are with the State Key Laboratory of Virtual Reality Technology and Systems, and Beijing Advanced Innovation Center for Biomedical Engineering, Beihang University, Beijing 100191, China, and also with the Research Unit of Virtual Body and Virtual Surgery (2019RU004), Chinese Academy of Medical Sciences, Beijing 100050, China.

\IEEEcompsocthanksitem J. Dai, S. Li and A. Hao are with Peng Cheng Laboratory, Shenzhen 518066, China. 
\IEEEcompsocthanksitem S. Li is also with Zhongguancun Laboratory, Beijing 100094, China.
\IEEEcompsocthanksitem H. Qin is with Department of Computer Science, Stony Brook University (SUNY at Stony Brook), Stony Brook, New York 11794-2424, USA.}

\thanks{Corresponding author: ham@buaa.edu.cn, qin@cs.stonybrook.edu.}}

%
%

\markboth{Journal of \LaTeX\ Class Files,~Vol. , No.~, 2023}%
{
}
%




\IEEEtitleabstractindextext{%
\begin{abstract}

High-accuracy, high-efficiency physics-based fluid-solid interaction is essential for reality modeling and computer animation in online games or real-time Virtual Reality (VR) systems. However, the large-scale simulation of incompressible fluid and its interaction with the surrounding solid environment is either time-consuming or suffering from the reduced time/space resolution due to the complicated iterative nature pertinent to numerical computations of involved Partial Differential Equations (PDEs). In recent years, we have witnessed significant growth in exploring a different, alternative data-driven approach to addressing some of the existing technical challenges in conventional model-centric graphics and animation methods. This paper showcases some of our exploratory efforts in this direction. One technical concern of our research is to address the central key challenge of how to best construct the numerical solver effectively and how to best integrate spatiotemporal/dimensional neural networks with the available MPM's pressure solvers. In particular, we devise the MPMNet, a hybrid data-driven framework supporting the popular and powerful Material Point Method (MPM), to combine the comprehensive properties of MPM in numerically handling physical behaviors ranging from fluid to deformable solids and the high efficiency of data-driven models. At the architectural level, our MPMNet comprises three primary components: A data processing module to describe the physical properties by way of the input fields; A deep neural network group to learn the spatiotemporal features; And an iterative refinement process to continue to reduce possible numerical errors. The goal of these special \textcolor{black}{technical developments} is to aim at involved numerical acceleration while preserving physical accuracy, realizing efficient and accurate fluid-solid interactions in a data-driven fashion. The extensive experimental results verify that our MPMNet can tremendously speed up the computation compared with the popular numerical methods as the complexity of interaction scenes increases while better retaining the numerical accuracy.

\end{abstract}

}

\maketitle

\IEEEdisplaynontitleabstractindextext

%
\IEEEpeerreviewmaketitle

\ifCLASSOPTIONcompsoc
\IEEEraisesectionheading{\section{Introduction and Motivation}\label{sec:introduction}}
\else
\section{Introduction and Motivation}
\label{sec:introduction}
\fi

%
%
%
%
\IEEEPARstart{D}{ynamic} fluid-solid coupling remains a popular
research direction among physics-based simulations and their
applications in computer graphics. The relevant phenomena are quite
common in our daily lives. 
\textcolor{black}{Famous physics-based methods include particle-based method~\cite{Koschier:2022:ASS}, grid-based method~\cite{chentanez:2011:RTE}, hybrid grid/particle method~\cite{Zhu:2005:ASF,Gao:2018:AnEF,Gao:2017:ANF,jiang:2015:APIC,Gao:2021:SMG}. Among them, Material Point Method (MPM)~\cite{Jiang:2016:MPM}, a hybrid grid/particle method, attracts much attention in recent years since it has been successful in simulating many fluid-like materials,} such as viscous
liquid~\cite{Ram:2015:AMP}, incompressible
liquid~\cite{Gagniere:2020:AHL,Fang:2020:IQM},
deformable~\cite{Fang:2019:SR} and fractured
materials~\cite{Wang:2019:SVD,Wolper:2020:AnisoMPM,Wolper:2019:CDM},
with more complex and convincing fluid-solid
dynamics.

Dynamic fluid-solid coupling is much more complex compared to single fluid modeling because of the additional, expensive calculation for handling and coupling with solid motion evolution and interaction. The tightly coupled physical modeling of fluid and solid dynamics, the complicated handling for the geometric and topological, and force-centric computational issues pertinent to the delicate boundary processing and coupling of the interface bifurcation are still bottlenecks that severely restrict the simulation efficiency and scene scalability. Benefiting from the increase of model complexity and numerical computation, MPM can simulate fluid and solid behaviors simultaneously in a unified framework. Nevertheless, the numerical computations will be time-consuming due to the complicated iterative calculations of the large-scale system of linear equations to arrive at an acceptable numerical convergence, such as Preconditioned Conjugate Gradient (PCG) method~\cite{Richard:1994:TSL,Golub:1996:MCE,Fedkiw:2001:VSS},
Incomplete Cholesky Preconditioned Conjugate Gradient (ICPCG) method~\cite{Foster:2001:PAL}, Gauss-Seidel (G-S) method~\cite{Stam:2003:RFD}, etc.

In contrast, data-driven methods have significantly progressed in recent years because of hardware improvement and model enhancement. Physical simulation can benefit from data-driven models by the demonstrated advantages such as fast extraction of parameters, accelerated convergence of calculation, fitting calculation process, enrichment of detail expression, etc. Researchers have proposed some excellent approaches to predict fluid behaviors. Famous data-driven methods include regression forests~\cite{Ladicky:2015:DFS}, Fully Connected Neural Networks (FCNN)~\cite{Luo:2016:DPM,Um:2018:LSM,Gao:2020:ALS}, Convolutional
Neural Networks
(CNN)~\cite{Tompson:2017:AEF,Xiao:2020:NCP,chen:2023:THA,Tumanov:2021:DPL},
Continuous Convolutional Neural Networks~\cite{Hiremath:2020:LFS,prantl:2022:GCM},
Recurrent Neural Networks (RNN)~\cite{Wiewel:2019:LSP,Wiewel:2020:LSS},
Graph Neural Networks
(GNN)~\cite{Li:2019:LPD,Sanchez-Gonzalez:2020:LSC}, Generated
Adversarial Networks (GAN)~\cite{Xie:2018:TGAN,cao:2022:BSM}, etc.
However, little attention has been paid to tackling fluid and solid's
two-way interactions since \textcolor{black}{different materials'} properties
and interacting forces are coupled in a complicated dynamic system. It
is still challenging to extract physics-aware features and laws
precisely. Thus, exploring a more efficient data-driven framework to
speed up numerical calculations while retaining physical correctness,
promises a significant research attempt.

In this paper, we aim to deeply and seamlessly combine data-driven and
model-driven methods to achieve efficient and accurate fluid-solid
coupling, with a specific goal of tackling the challenging yet
intriguing interactive pressure computations. In particular, we design
new neural networks to predict fluid dynamics over time, which can
enhance the computing power w.r.t. interaction and motion towards
fluid-solid coupling. Our fluid-solid coupling
prediction is built upon Interface Quadrature Material Point Method (IQ-MPM)'s numerical
solvers~\cite{Fang:2020:IQM}, inherits the advantages of grid-particle
solvers that handle advection with accuracy and maintain
incompressibility conditions. To break the bottleneck that restricts the time efficiency
of the MPM method, we combine MPM and neural networks using a
collaborative end-to-end structure for accelerated pressure
solving. As a viable technical solution, three different neural
network components will correspond to the decoupled specific physical
processes: fluid dynamics, solid motion, and surface
interaction. The architectural pipeline comprises an encoder,
a Convolutional Long Short-Term Memory (ConvLSTM), and a decoder for spatial
compression and temporal prediction. Depending on the simulation scales
and accuracy requirement, the data-driven procedure can afford the
predicted pressures that get rid of dozens to even hundreds of
time-consuming iterations. Instead, the numerical solver takes over
the predicted pressures and fast convergence can be achieved with only
a few iterations. The deep and seamless integration of physical
simulation and the machine learning process takes advantage of each,
which could yield significantly faster than traditional numerical
simulators while retaining physical accuracy. In summary, the salient
contributions of this paper include:
\begin{itemize}
  
\item
We advocate a hybrid data-driven approach which efficiently combines
the end-to-end network architecture with MPM to speed up numerical
computation, while guaranteeing physical reality.

\item
We detail a matrix representation method towards different materials'
pressure characteristics and their interface to efficiently represent the 3D spatial coupling and interaction in the mixing scenes involving both fluid and solid. 
 And we exploit optimized training techniques with Huber loss function to
further improve prediction robustness. 

\item
We introduce a new parameter $\zeta$ to quantitatively describe the interacting complexity level for fluid-solid interaction during animation. As this parameter increases, our proposed method yields a significant numerical speedup. The best performance achieved in our extensive experiments is 28$\times$ faster than the numerical solutions.

\end{itemize}

\section{Background and Related Works}


In this section, we have reviewed the prior works related to material point method, data-driven-based fluid modeling, and fluid-solid coupling modeling.

\subsection{Material Point Method}
As a hybrid Lagrangian/Eulerian discretization method, the MPM technique extends the FLuids-Implicit-Particle (FLIP) method from Computational Fluid Dynamics (CFDs) to solid mechanics. Since its introduction in 1994 by Sulsky et al.~\cite{sulsky:1994:PMH,sulsky:1995:APM}, it has proven to be very effective in computer graphics for simulating various solid and fluid materials. Stomakhin et al.~\cite{stomakhin:2013:MSS} proposed a constitutive model that can deal with various dense and wet snow behaviors. Hu et al.~\cite{hu:2018:MLS} presented the Moving Least Squares Material Point Method (MLS-MPM) that can handle coupling with rigid bodies and cutting, which MPM did not previously support. Fang et al.~\cite{Fang:2020:IQM} proposed an IQ-MPM algorithm for strong two-way coupling of incompressible fluids with volumetric elastic solids. Su et al.~\cite{su:2021:USA} extended MPM for simulating various viscoelastic liquids with phase change. Li et al.~\cite{li:2021:BFE} proposed BFEMP, which monolithically coupling the MPM with the Finite Element Method (FEM) for elastodynamics through frictional contact. Moreover, even thin elastic membranes and shells~\cite{guo:2018:TSF}, wet clothing~\cite{fei:2018:MMS}, fracture animation~\cite{Wolper:2019:CDM} have been successfully applied by MPM.

Compared to other numerical methods, MPM has a more complex constitutive model and numerical solver, enabling it to simulate various physical dynamics. However, the iterative solution of the pressure matrix is the guarantee of simulation accuracy yet the bottleneck of calculation efficiency as well. It is of great significance to find a new method to solve the pressure quickly while ensuring accuracy.

\subsection{Data-driven Fluid Modeling}
Traditional high-resolution fluid simulation methods always consume substantial computational resources. 
Pressure calculation is the most time-consuming and challenging part of fluid simulation. Although the Laplacian matrix is symmetric and positive semidefinite, linear systems usually have many free parameters, which means that standard iterative solvers must carry out a large number of iterations to produce as little error as possible. The count of iterations is heavily dependent on the data size. 

With the rapid improvement of neural network inference performance, many data-driven models have been proposed~\cite{Ladicky:2015:DFS,Um:2018:LSM,Peng:2019:DDR}. \textcolor{black}{Tang et al.~\cite{tang:2022:NGF} regressed a Green’s function solution for 2D Laplacian systems aided by the fully-connected networks.} Yang et al.~\cite{Yang:2016:DDP} used a fully connected network to replace the local solution of PCG for accelerating 3D smoke simulation. The calculation speedup of the pressure mapping method is more than ten times faster than that of the traditional manner without multi-threaded calculation. 
In addition, Gao et al.~\cite{Gao:2020:ALS} improved the above method to make it more suitable for liquid simulations. They fed the liquids' level set and velocity properties into networks to consider the differences between smoke and liquids. What is more, Xiao et al.~\cite{Xiao:2020:NCP} used CNN to solve the pressure Poisson equation globally, which is suitable for large-scale scenes. Tompson et al.~\cite{Tompson:2017:AEF} fed the gradient of the intermediate velocity into CNN to replace the Euler pressure projection, and then the velocity gradient was predicted. The runtime and accuracy are superior to Jacobi's method, and its visible results are comparable to PCG's while being orders of magnitude faster. Kim et al.~\cite{kim:2019:DFG} presented a new generative network to accelerate synthetic fluid simulation from a set of reduced parameters. Wiewel et al.~\cite{Wiewel:2020:LSS} divided the input velocity and density matrix into two parts of the latent spatial domain and conducted time dimensional prediction using a subdivided encoder and LSTM, respectively, which yields significant speedups compared to traditional solvers in time predictions for smokes. Instead of numerical acceleration, detail enhancement is another way to improve the modeling efficiency and accuracy. Um et al.~\cite{um:2020:SLL} achieved significant reductions in numerical errors in PDE-solvers by training artificial neural networks and differentiable physics solvers.
Chu et al.~\cite{Chu:2017:DDS} used twin CNN to judge whether two high and low-resolution patches in the database are the same parts of the fluid. Then the high-resolution patch is synthesized into the low-resolution image to improve the resolution. Xie et al.~\cite{Xie:2018:TGAN} first used GAN for four-dimensional functions to train the mapping relationship from low-resolution to high resolution for getting higher resolution. Xiao et al.~\cite{Xiao:2019:CFC} presented a new fast CNN-based shape correction method, which can simulate preview at low-resolution but maintain high-resolution fluid shape. However, because the network input is local information, it cannot guarantee the divergence-free condition. To this end, Li et al.~\cite{Li:2020:LPP} proposed a novel CNN-based regression model to estimate the physical parameters of Eulerian gas, and the learned parameters are used to guide the high-resolution simulations by combining the CNN-based velocity regression.

To conclude, many state-of-the-art data-driven works have emerged in recent years for fluid-related modeling and achieving convincing performances. Whereas the interaction of fluid and dynamic solids have few involved. Because the \textcolor{black}{different properties} and interacting forces of fluid and solid are coupled in a complicated interacting scene, dynamic behaviors prediction with a uniform network for all materials may result in insufficient details and artifacts.

\subsection{Data-driven Fluid-solid Coupling Modeling}
Although there are many well-established numerical solutions for fluid-solid coupling simulation, such as Lagrangian solids coupling to Eulerian fluids~\cite{robinson:2011:SPD}, Lagrangian solids coupling to Lagrangian fluids~\cite{Gissler:2019:ISP}, Euler-based solids and liquid~\cite{Teng:2016:ESF}, hybrid particle-grid solvers~\cite{daviet:2016:SIM,Fang:2020:IQM}
, it is not an easy task to predict the two-way coupled behaviors of solids and fluids by a data-driven framework.

Takahashi et al.~\cite{Takahashi:2021:DFS} presented a differentiable one-way coupling of fluid with a rigid objects simulator using deep neural networks to learn dynamics and solve control problems. It is typically necessary to compute the fluid and solid dynamics using their respective motion and interaction calculations. In addition, the two-way fluid-solid coupling will pose additional challenges because of the increased solid interaction, which further makes it difficult to learn and predict fluid dynamics and solid motion. Sanchez-Gonzalez et al.~\cite{Sanchez-Gonzalez:2020:LSC} proposed a Lagrangian-based way using interactive GNN to rapidly propagate the effects between adjacent nodes in space dimensions. It has better dynamic characteristics between fluids and solids but is limited by particles number with physical understanding, which can only predict scenarios with thousands or tens of thousands of particles. Li et al.~\cite{li:2020:VGL} used an intermediate representation of particles, which for dealing with solid bodies, deformable objects, and fluids. The input simulation picture is used to predict the rough position and group properties of the particle through the visual prior neural network but is still limited by the number of particles.

In this direction, deeply combining the data-driven and physics-based models, recovering the complex physical processes without distortion yet significantly improving the calculation speed is a challenging and inspiring attempt.

\section{Method Overview}

\begin{figure*}[htb]
\centering
\setlength{\abovecaptionskip}{-0.2cm} 
\includegraphics[width=1\textwidth]{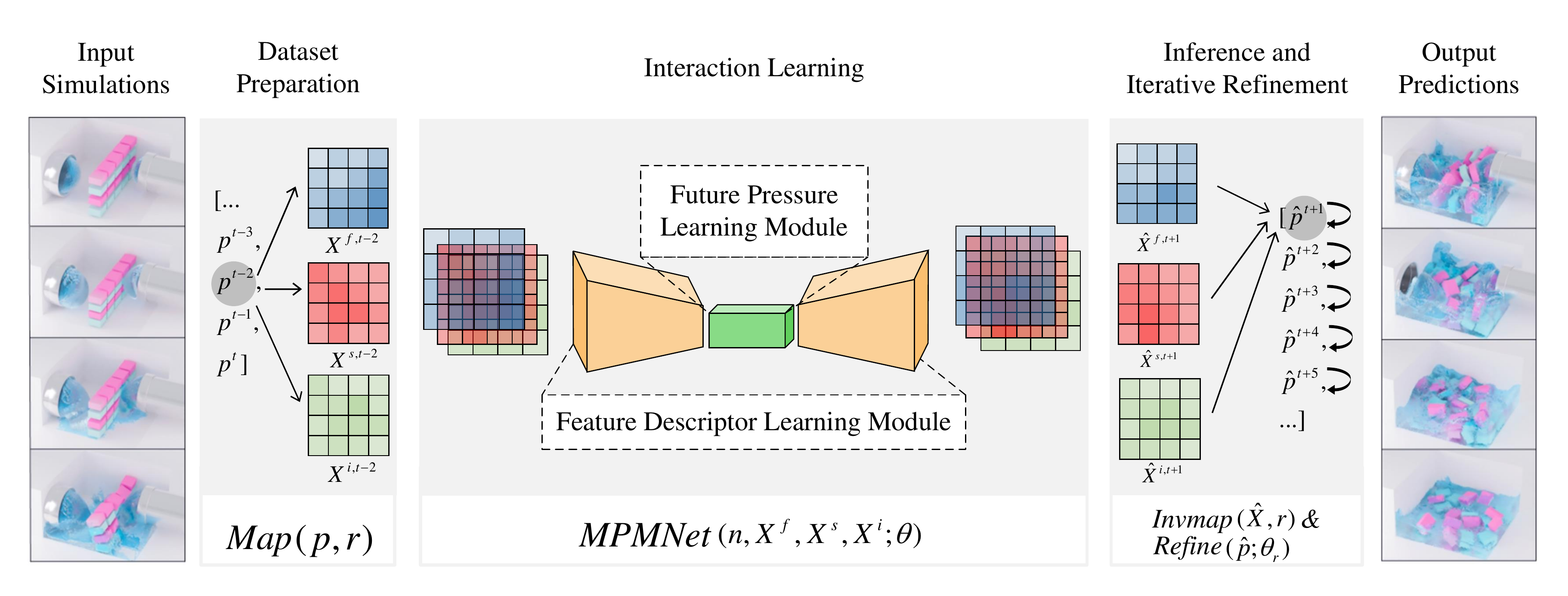}
\caption{The workflow of our MPMNet framework. The input data used during the training phase is prepared using physics-based solvers and processed by mapping to three Cartesian grids. Our network encodes and learns temporal inference using pressure grids. These predicted fields are then refined and applied directly to the simulations for fast prediction.}
\label{fig:pipeline}
\end{figure*}

\subsection{Basic MPM Model}

\textcolor{black}{We focus on fluid-solid coupling simulation, for which we employ the well-established IQ-MPM model, where Affine Particle-In-Cell (APIC) method is used to transfer grid quantities.} It has the advantages of efficiently handling self-collision, single monolithic solutions for the coupled pressure fields, and supporting of multi-material interactions.
To make our work more self-contained, we start with a brief review of the pressure solution in the MPM model. The single material governing equations~\cite{Bonet:2008:NCM} are:
\begin{equation}
\left\{
\begin{alignedat}{2}
\frac{D \rho^k}{D t}+\rho^k\nabla\cdot \textbf{\emph{v}}^k=0,&\qquad \textbf{\emph{x}}\in\Omega^k\qquad\quad\\
\rho^k\frac{D \textbf{\emph{v}}^k}{D t}-\nabla\cdot\sigma^k-\rho^k\textbf{\emph{g}}=0,&\qquad \textbf{\emph{x}}\in\Omega^k\qquad\quad
\end{alignedat}
\right.
\end{equation}

\begin{equation}
\left\{
\begin{alignedat}{2}
&\sigma^k\cdot \textbf{\emph{n}}^k=\textbf{\emph{u}}^k_N,\qquad\quad& \textbf{\emph{x}}\in\partial\Omega^k_N\\
&\textbf{\emph{v}}^k\cdot \textbf{\emph{n}}^k=\textbf{\emph{u}}^k_S,\qquad\quad & \textbf{\emph{x}}\in\partial\Omega^k_S\\
&\textbf{\emph{v}}^k=\textbf{\emph{u}}^k_{NS},& \textbf{\emph{x}}\in\partial\Omega^k_{NS}
\end{alignedat}
\right.
\end{equation}

\begin{equation}
\left\{
\begin{alignedat}{2}
&(\textbf{\emph{v}}^s-\textbf{\emph{v}}^f)\cdot \textbf{\emph{n}}^s=0,\qquad\quad& \textbf{\emph{x}}\in\Gamma\label{equ:slipv}\\
&p^s-p^f=0,   &\textbf{\emph{x}}\in\Gamma,
\end{alignedat}
\right.
\end{equation}  
where $k\in\{f,s\}$ is a label denoting the fluid domain $\Omega^f$ or the solid domain $\Omega^s$. $\rho^k$, $\textbf{\emph{v}}^k$, $\sigma^k$, $\textbf{\emph{n}}^k$, $p^k$ denote density, velocity, the Cauchy stress, outward pointing normal, and the pressure of $\Omega^k$, respectively. \textcolor{black}{The gravity force is denoted by $\textbf{\emph{g}}$.} $\textbf{\emph{u}}^k_N$, $\textbf{\emph{u}}^k_S$, $\textbf{\emph{u}}^k_{NS}$ represent the free surface conditions in domain $\Omega^k_{N}$, slip boundary condition in domain $\Omega^k_{S}$, and no-slip boundary condition in domain $\Omega^k_{NS}$, separately. $\boldsymbol{\Gamma}$ denotes the interface between domain $\Omega^f$ and $\Omega^s$. Equation~\ref{equ:slipv} enforces the normal velocity continuity and pressure continuity to realize the free slip boundary condition. It should be noted that $\nabla\cdot \textbf{\emph{v}}^k=0$ is to ensure the material's incompressibility.

 On this basis, Fang et al.~\cite{Fang:2020:IQM} proposed IQ-MPM that first simulates stable two-way coupled interactions between nonlinear elastic solids and incompressible fluids with large time steps without requiring multiple monolithic solves.
 Its pressure-only system is given by the large linear equations, which are solved iteratively:
\begin{equation}
\setlength{\arraycolsep}{2.0pt}
\left( \begin{array}{cccc}
A_{11}&0&0&A_{14}\\
0&A_{22}&A_{23}&A_{24}\\
0&A_{23}^T&A_{33}&A_{34}\\
A_{14}^T&A_{24}^T&A_{34}^T&A_{44}
\end{array} \right )\!\!\!
\setlength{\arraycolsep}{0.5pt}
\left ( \begin{array}{cccc}
\textbf{\emph{p}}^{s,t+1}\\
\textbf{\emph{p}}^{f,t+1}\\
\textbf{\emph{y}}^{t+1}\\
\textbf{\emph{h}}^{t+1}
\end{array} \right )
\mathclap{=}
\setlength{\arraycolsep}{0.5pt}
\left ( \begin{array}{cccc}
\frac{S^s \textbf{\emph{p}}^{s,t}}{\Delta t}-G^{s^T} \textbf{\emph{v}}^{s,t}\\
G^{f^T} \textbf{\emph{v}}^{f,t}\\
B\textbf{\emph{v}}^{f,t}-\textbf{\emph{b}}\\
H^{s}\textbf{\emph{v}}^{s,t}-H^{f}\textbf{\emph{v}}^{f,t}
\end{array} \right )\mathclap{,}
\label{equ:solve}
\end{equation}
where
\begin{equation}
\left\{
\begin{alignedat}{2}
& A_{11}=\frac{S^s}{\Delta t}+\Delta tG^{s^T}M^{s^{-1}} G^s,\\& A_{14}=-\Delta t G^{s^T}M^{s^{-1}}H^{s^T},\\&A_{22}=\Delta tG^{f^T}M^{f^{-1}}G^f,\\& A_{23}=\Delta tG^{f^T}M^{f^{-1}}B^T\\&A_{24}=\Delta tG^{f^T}M^{f^{-1}}H^{f^T}, \\&A_{33}=\Delta tBM^{f^{-1}}B^T,\\&A_{34}=\Delta tBM^{f^{-1}}H^{f^T}, \\&A_{44}=\Delta t(H^sM^{s^{-1}}H^{s^T}+H^fM^{f^{-1}}H^{f^T}).
 \end{alignedat}
 \right.
\end{equation}
$S^s$ is the scaling matrix for solid pressure, $G^s$ is the solid domain gradient operator, $M^f$, $G^f$, $B$ refer to the lumped mass, gradient, and boundary operators in the fluid domain $\Omega^f$, separately. $\textbf{\emph{b}}$ describes boundary condition. $H^s$ and $H^f$ denote the solid and fluid coupling terms, respectively. The most important properties are fluid pressure $\textbf{\emph{p}}^{f,t+1}$, solid pressure $\textbf{\emph{p}}^{s,t+1}$, slip boundary pressure $\textbf{\emph{y}}^{t+1}$, and solid-fluid interface pressure $\textbf{\emph{h}}^{t+1}$ that need to be iteratively solved. More complicity of the model, or higher convergence accuracy, means more numerical computing burden and more time-consuming. This is where we seek to introduce data-driven models to break through the bottleneck.

\subsection{Method Pipeline}

Our MPMNet architecture is inspired by Latent Space Physics method~\cite{Wiewel:2019:LSP}, which encodes pressure of single-phase fluid into a latent space and utilizes a LSTM to predict the changes of pressure fields over time. Furthermore, our approach broadens the scope of data-driven fluid prediction and speeds up the computational process. The characteristics of fluid-solid two-way interactions that differ from single fluid motion require the development of a novel data processing technology. We \textcolor{black}{separate} the fluid and solid data, mapping it into three distinct representations before feeding it into the network via separate channels. Pressure is transferred longitudinally between network layers to account for fluid and solid motion. While the forces are transferred transversely between channels to learn the interactive behaviors. In addition, to account for time dimension evolution, we utilize a ConvLSTM to update the spatiotemporal information. Our dataset comprises plenty of fluid-solid coupling scenes generated by IQ-MPM~\cite{Fang:2020:IQM}. The complete solution can be summarised into three main components: (as shown in Fig.~\ref{fig:pipeline}). 

\textbf{Dataset Preparation.} The dataset is built in three stages: (1) generating fluid-solid coupling samples as our training set to encode coupling behaviors (i.e., dam-breaking scenes, solids falling into the water scenes, etc.) using different subsets of initial conditions with several consecutive frames, (2) constructing the pressure fields of fluid and slip boundary, the solid, the interface between fluid and solid bodies, and the nodes' coordinate of the Cartesian grids corresponding to each value, and (3) mapping the fields into the corresponding Cartesian grids to obtain a suitable format for more compatibility with the neural network.
 
\textbf{Interaction Learning.}  
We use a collaborative end-to-end neural network to learn and encode interactions between fluid and solid bodies. 
By feeding the pressure properties at the beginning of a few steps to our trained end-to-end network, we can predict the pressure in many of the following steps that are dozens of times the amount of the initial input steps.

\textbf{Inference and Iterative Refinement.} 
We utilize a post-processing step to refine the pressure of fluid and solid while reducing the inference noise to ensure the accuracy and robustness of MPMNet. Meanwhile, an improved Huber loss function and Algebraic Multigrid-preconditioned
(AMG)~\cite{demidov:2019:AMG} or Gauss-Seidel (G-S)~\cite{yoon:1988:LSG} based refinement is performed to ensure the precision and incompressibility of our hybrid model.
In each step, instead of numerous iterated computations, we utilize the inferred pressure obtained from our learning pipeline to be mapped back to the numerical solver. We only need a few iterations to meet a satisfactory convergence accuracy based on the predicted pressures. Through such deep integration of data-driven and physical-driven models, MPMNet can significantly reduce iterative computations. 

Algorithm~\ref{Prediction Algorithm} outlines the details of our MPMNet inference scheme.
\begin{algorithm}
    \begin{algorithmic}[1]
      \For {$t$ in $n$}
          \State particle-to-grid transfer
          \State coupled solve (equation~\ref{equ:solve})
          \State get $\textbf{\emph{p}}^{f,t}$, $\textbf{\emph{p}}^{s,t}$, $\textbf{\emph{y}}^{t}$,$\textbf{\emph{h}}^{t}$ and the corresponding node coordinate $\textbf{\emph{r}}$ from physical method
          \State map pressure fields to effective matrices $X^{f,t}$, $X^{s,t}$, $X^{i,t}$; ($X=Map(\textbf{\emph{p, r}})$) 
          \State normalization procedure (see Sec 4.1)
          \State grid-to-particle transfer
      \EndFor
      \State get MPMNet input $I=[n,X^f,X^s,X^i]$ 
      \For {$t$ in $m$}
          \State particle-to-grid transfer
          \State MPMNet inference and output $\hat{X}^{f,t}$, $\hat{X}^{s,t}$, $\hat{X}^{i,t}$
          \State map effective matrices to pressure fields $\hat{\textbf{\emph{p}}}^{f,t}$, $\hat{\textbf{\emph{p}}}^{s,t}$, $\hat{\textbf{\emph{y}}}^{t}$, $\hat{\textbf{\emph{h}}}^{t}$; ($\hat{\textbf{\emph{p}}}=Invmap(\hat{X},\textbf{\emph{r}})$)
          \State Iterative refinement (see Sec 4.4)
          \State grid-to-particle transfer
      \EndFor
    \end{algorithmic}
    \caption{Pseudo-code for data processing and MPMNet inference}
    \label{Prediction Algorithm}
  \end{algorithm}

\section{MPMNet Model}

\subsection{Data Processing}
\begin{figure}[htb]
\centering
\setlength{\abovecaptionskip}{0cm} 
\includegraphics[width=0.49\textwidth]{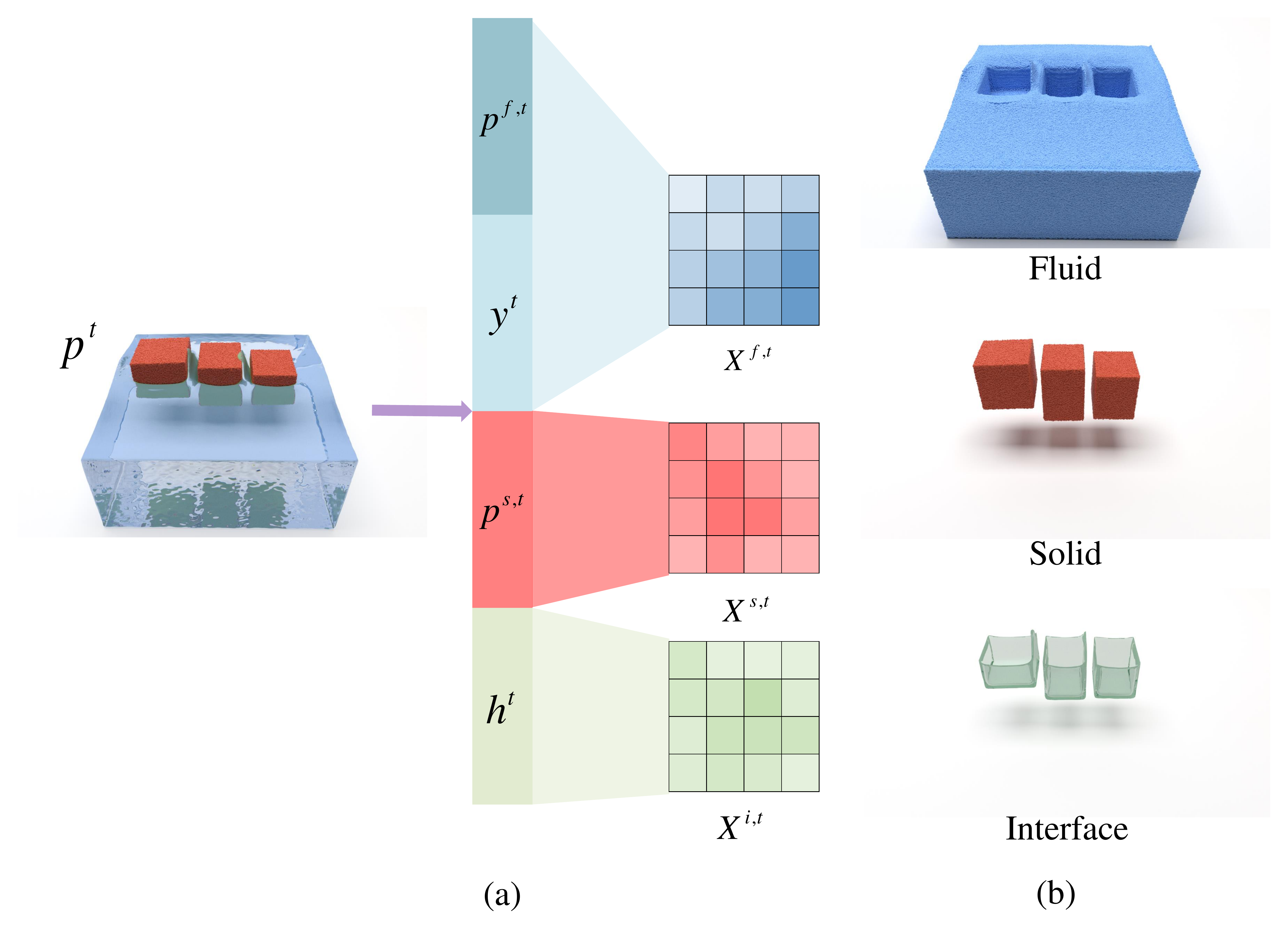}
\caption{Pressures projection. (a): Pressure fields are independently mapped to three Cartesian grids to better understand the physical significance; (b): The three separated representations are the fluid, solid, and interface, from top to bottom.}
\label{fig:map}
\end{figure}

Efficient data extraction and accurate mapping of physical features are \textcolor{black}{key} to data processing. Of which, the main goal is to compute the decoupled pressure matrices in consecutive frames from the MPM solver. \textcolor{black}{The pressure in our method is obtained from all grid nodes with non-zero mass, as these nodes contribute to the compression of the sparse pressure matrices in the physical solver. In our implementation, each value in the pressure field corresponds to the coordinates of the respective grid node. We construct pressure matrices only for the Cartesian grid, where each value corresponds to the coordinates of its respective grid node.} However, if we feed the fields directly into the neural network, spatial information will be difficult to learn. On the one hand, as each step updates, the total count of non-zero-mass nodes in each step changes, so the length of the pressure fields computed in each frame is unfixed, which cannot be fed into the CNN as a structured form. On the other hand, each value represents the pressure at its corresponding grid node. Its absolute position in the field has no relation to its value change. If the corresponding node coordinate is ignored, it is difficult for neural networks to learn. Thus, as shown in Fig.~\ref{fig:map} (a), we map these pressure fields to three Cartesian grids separately for better learning their physical behaviors. Fluid pressure $\textbf{\emph{p}}^{f,t}$ and slip boundary pressure $\textbf{\emph{y}}^{t}$ are mapped to the first grid $X^{f,t}$ to represent fluid pressure over time, solid pressure $\textbf{\emph{p}}^{s,t}$ is mapped to the second grid $X^{s,t}$ to express the pressure on the solid over time, and the pressure at the fluid-solid interface $\textbf{\emph{h}}^{t}$ is mapped to the third grid $X^{i,t}$ to record the pressure change at the interactive surface. More explicitly, Fig.~\ref{fig:map} (b) shows the three separated representations for a scene at $t=200$, from top to bottom, as the fluid, the solid, and the interface. To sum up, we have a general equation $X=Map(\textbf{\emph{p}},\textbf{\emph{r}})$ with $\textbf{\emph{p}}=Invmap(X,\textbf{\emph{r}})$ to get the effective matrices mapping relation for the neural network, where $\textbf{\emph{p}}$ denotes the value of pressure fields and $\textbf{\emph{r}}$ denotes its corresponding coordinate node.

\begin{figure}[htb]
\centering
\setlength{\abovecaptionskip}{0cm} 
\includegraphics[width=0.3\textwidth]{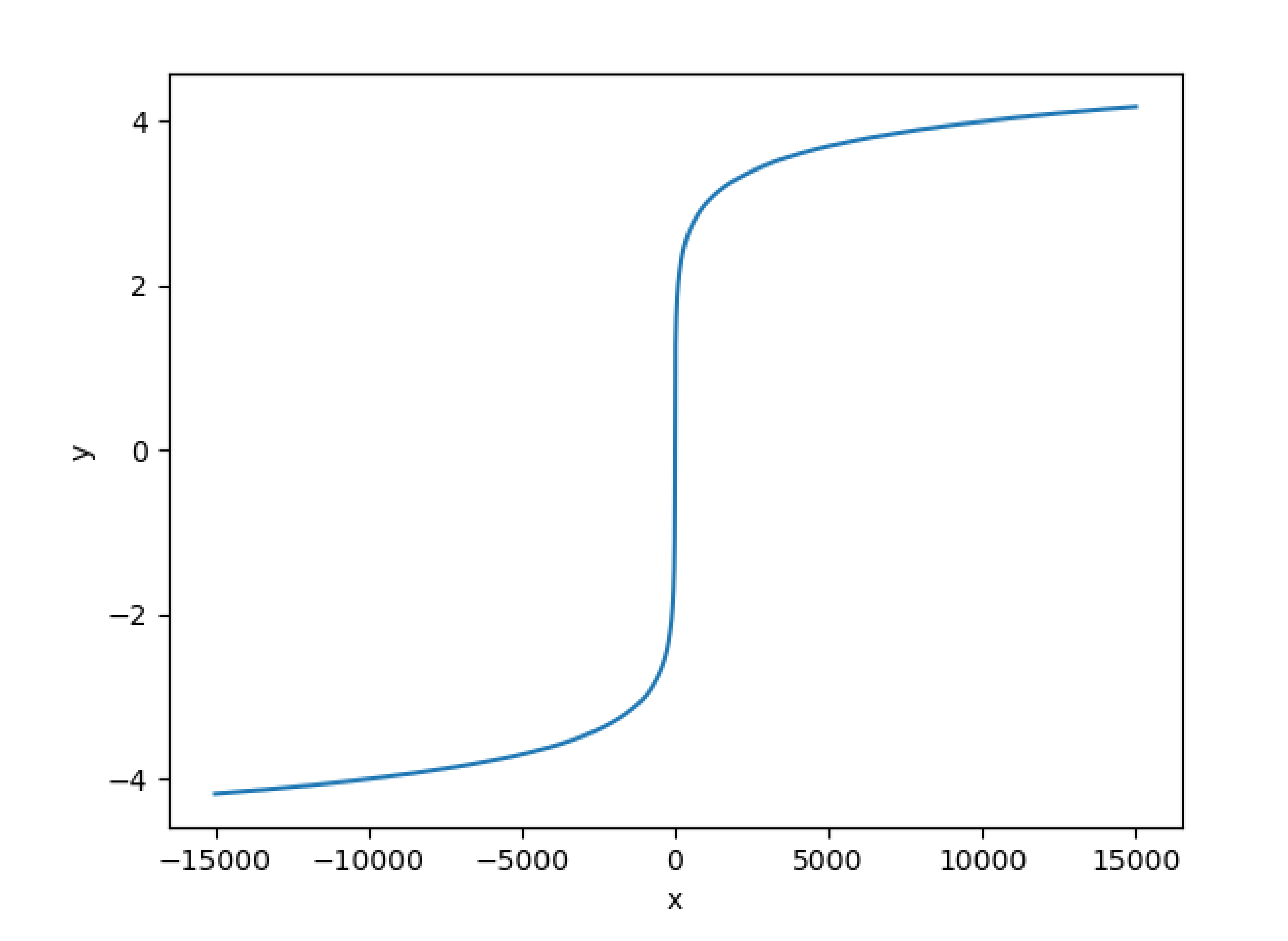}
\caption{The modified $log$ function for data normalization.}
\label{fig:log}
\end{figure}

In addition, in our training process, \textcolor{black}{gradient vanishing can occur} when using the backpropagation (BP) method~\cite{riedmiller:1993:BP} to train the neural network. To overcome this problem, we normalize the original dataset using the modified $\log$ function (Equation~\ref{equ:log}), as shown in Fig.~\ref{fig:log}. The normalization procedure enhances the stability of our MPMNet solver much further.

\begin{equation}
y=\left\{
\begin{alignedat}{2}
&\lg(x+1),\qquad\quad\quad  x\ge 0\\  
&-\lg(-x+1),\qquad  else
\end{alignedat}
\right.
\label{equ:log}
\end{equation}

We design and simulate plenty of different fluid-solid coupling scenes to generate the original training data, where the positions of solids are randomly generated. We create fluid blocks in the scenes to impact different solids (i.e., cubes, spheres, and bear models) and record the pressure matrices in 585 consecutive frames. Each scene we used has from 1 million to 10 million particles located in the domain of the fixed size. By simulating the interaction process between several fluids and solid bodies with different positions, densities, and shapes, we obtain the pressure values computed in each simulation step and map them to the corresponding grids to generate our dataset. In detail, 5 dam-breaking scenes (585 frames per scene), 5 solids-falling scenes (585 frames per scene), and 5 water-dropping scenes (585 frames per scene), were simulated at the fixed resolution of $32^3$.

\begin{figure}[htb]
\centering
\setlength{\abovecaptionskip}{0cm} 
\includegraphics[width=0.45\textwidth]{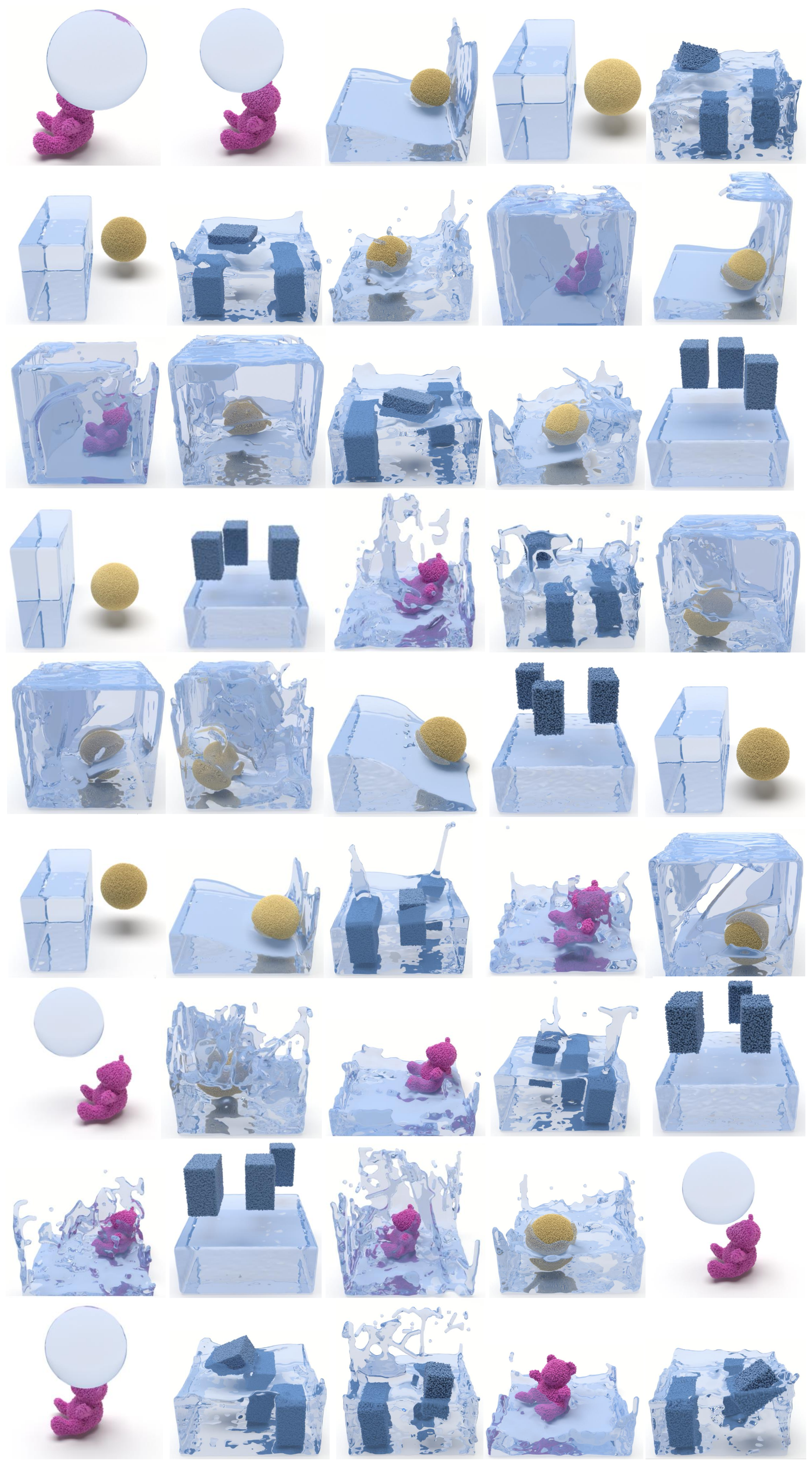}
\caption{An example of fluid-solid coupling dataset scenes.}
\label{fig:dataset}
\end{figure}

\subsection{Network Architecture}
Our proposed learning architecture, MPMNet, is a multi-channel model to successfully capture the inherent properties of fluid-solid coupling simulation. Its novel end-to-end learning architecture for quickly predicting the pressure properties of subsequent frames is divided into two main modules: the feature descriptor learning module and the future pressure learning module.


\subsubsection{Feature Descriptor Learning}

Our dataset involves empty grid nodes because the fluids and solids will not naturally occupy the entire fluid-solid coupling simulation scenarios. Hence, most pressure values are zero, which means our input data contains many sparse matrices. To this end, we implement an improved autoencoder in the feature descriptor learning module through unsupervised learning. After training the autoencoder, we can extract the feature descriptor with dimensions far smaller than the sparse input data. Along this direction, we learn the essential characteristics of our pressure matrices through this improved autoencoder to reduce data redundancy and increase information flow, which will be applied to future pressure prediction modules.

\begin{figure}[htb]
\centering
\setlength{\abovecaptionskip}{0cm} 
\includegraphics[width=0.49\textwidth]{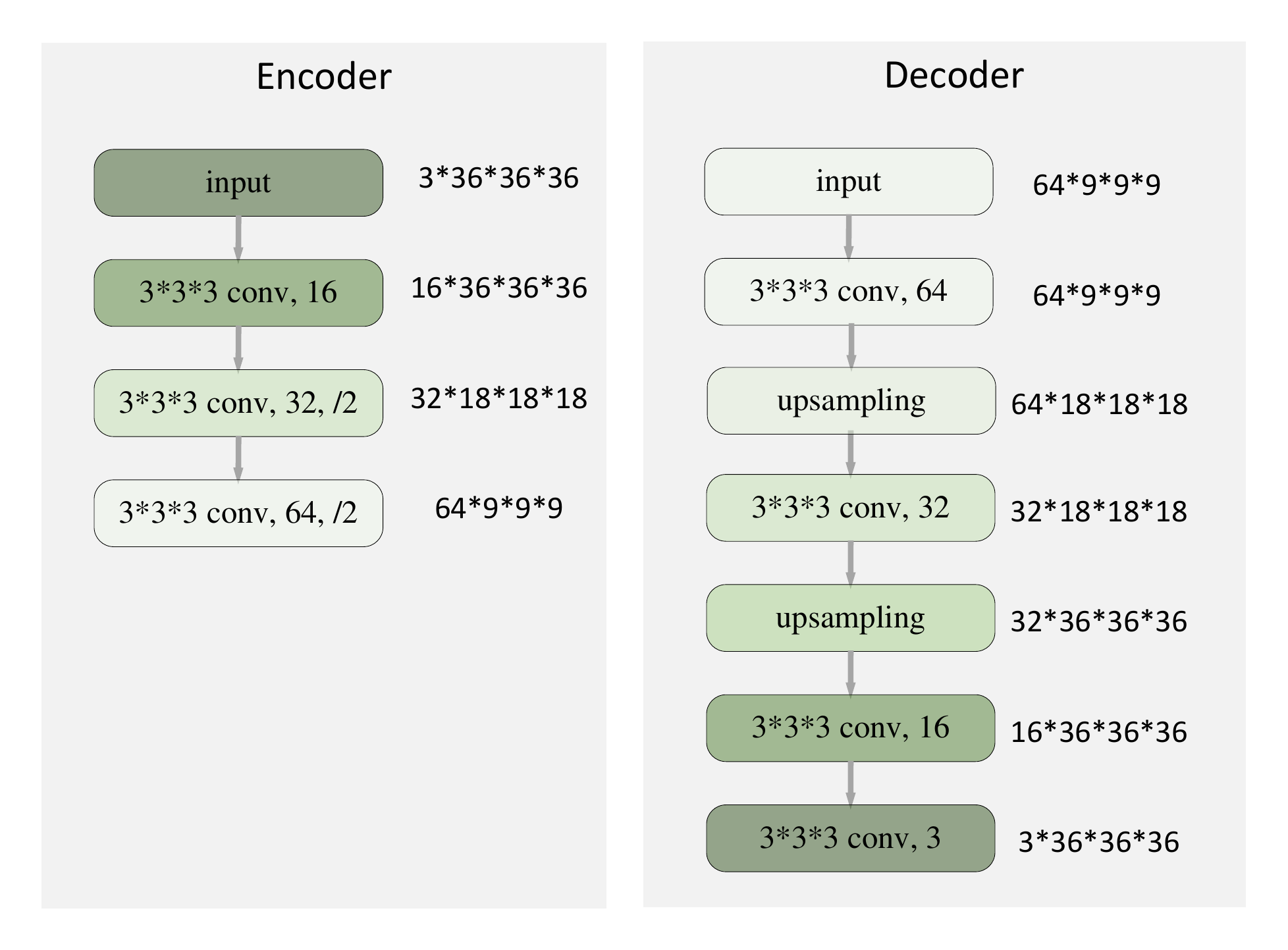}
\caption{The detailed network of our feature descriptor learning architecture, which includes an encoder and a decoder for feature learning.}
\label{fig:network}
\end{figure}

Fig.~\ref{fig:network} shows the detailed feature descriptor learning architecture. We implement three layers of 3D CNN in the encoder. Similarly, the decoder is formulated with four 3D CNN-based layers. The output of our encoder is described as $9^3$ feature matrices with 64 channels. We use LeakyRelu as the activation function because it is a good way to deal with the gradient vanishing problem~\cite{Hochreiter:2001:GFR}.

The trained autoencoder can learn the feature descriptors automatically. We present a large dataset that contains the spatiotemporal pressure of a fluid-solid coupling simulation that can be fed to our networks. We define the input training set as $I$, which is the pressure matrices in consecutive frames. With the given input field $I=[n,X^f,X^s,X^i]$, where $n$, $X^f\in \mathbb{R}^{d\times h\times w}$, $X^s\in \mathbb{R}^{d\times h\times w}$ and $X^i\in \mathbb{R}^{d\times h\times w}$ denote the count of frames inputted, the pressure of fluid and slip boundary, the pressure of solid and the pressure of the interface of fluid-solid in scenes, separately. Explicitly, we use $X^i$ to emphasize the change in the fluid-solid coupling interface. In our experiment, $d=h=w=36$. The network input is $36^3$ because of the four boundary pressure cells in each dimension when we train at $32^3$ models.
The input pressures $(X^{f,t},X^{s,t},X^{i,t})$ at time $t$ are embedded into a latent spatial domain through the encoder:
\begin{equation}
C^t=Enc(X^{f,t},X^{s,t},X^{i,t};\theta_e),
\end{equation}
where $Enc$ is the encoder function with LeakyRelu non-linearity, $\theta_e$ is the encoder network parameter, 
The matrix $C^t$ denotes the latent spatial feature information at time $t$, which is one input of the ConvLSTM network.

The decoder is broadly similar to the reverse process of the encoder, restoring data from latent space to a readable dimension using the physical model. The equation is defined as follows:
\begin{equation}
[\hat{X}^{f,t},\hat{X}^{s,t},\hat{X}^{i,t}]=Dec(C^t;\theta_d),
\end{equation}
where $\theta_d$ is the decoder network parameter and $\hat{X}$ is used to represent generated data from our decoder to distinguish the ground truth.

\subsubsection{Pressure Learning and Prediction}

To extract node features from past pressure values of all feature descriptors, we use the future pressure learning module to capture the temporal dependency among all the data of the feature descriptors. Since recurrent neural networks have inherent advantages for modeling sequential data, a ConvLSTM is implemented in the future pressure learning module for consecutive frame learning.

\begin{figure}[htb]
\centering
\setlength{\abovecaptionskip}{0cm} 
\includegraphics[width=0.49\textwidth]{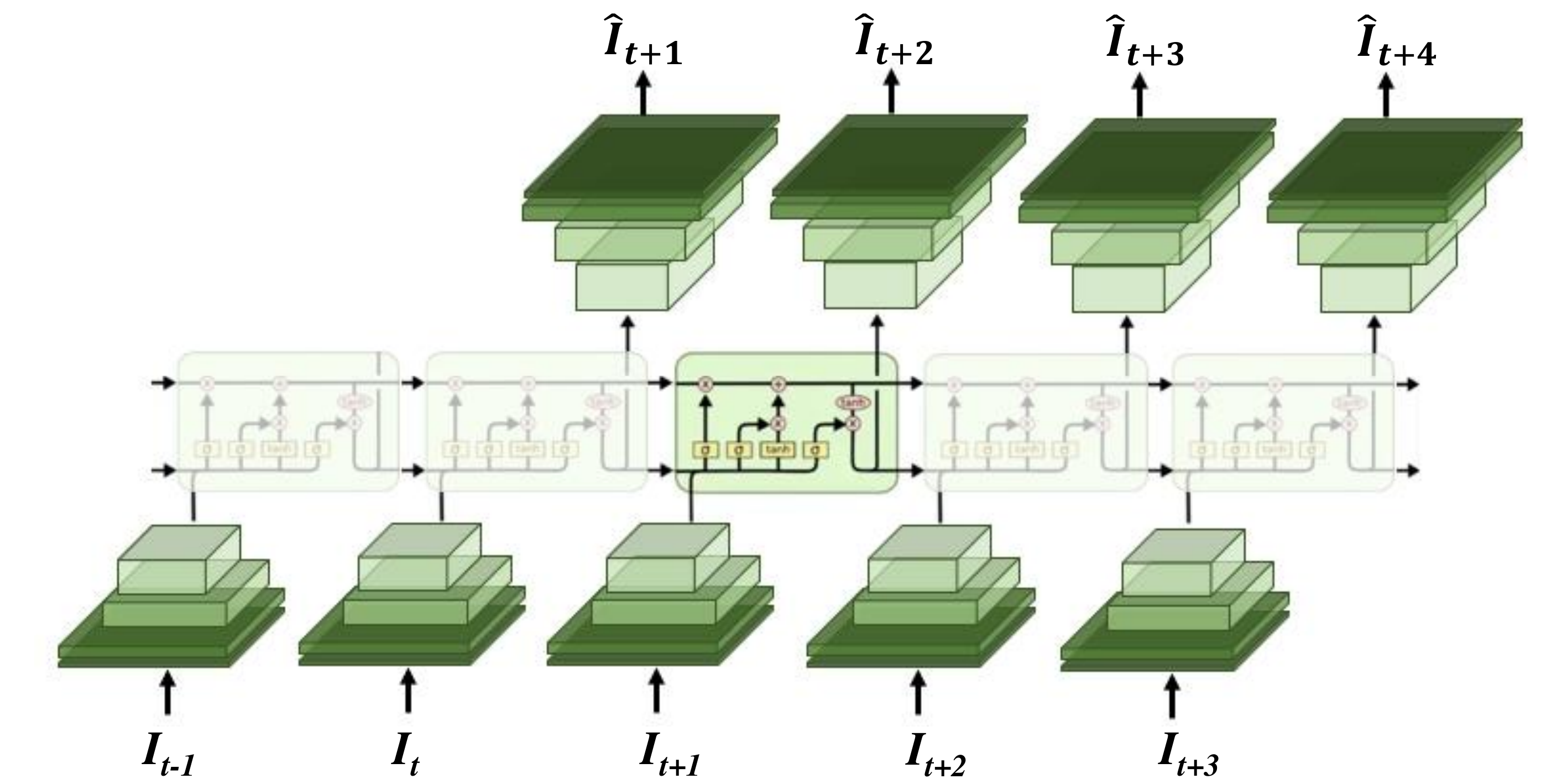}
\caption{A simplified example of the learning process of future pressure learning module.}
\label{fig:lstm}
\end{figure}

We design a future pressure learning module consisting of a ConvLSTM and a 3D CNN-based layer to obtain the following feature information of pressures from previous $n$ time steps.
The previous latent space $C^{(t-n+1)}$, $C^{(t-n+2)}$,..., $C^{t}$, from time $(t-n+1)$ to $t$, are extracted and fed into the future pressure learning module for integration. Assume the output of the future pressure learning module is $\hat{C}^{(t+1)}$, the $\hat{C}^{(t+1)}$ is computed as:
\begin{equation}
\hat{C}^{(t+1)}=Fpl(C^{(t-n+1)},C^{(t-n+2)},...,C^{t};\theta_f),
\end{equation}
where $Fpl$ represents the future pressure learning module, and $\theta_f$ is the corresponding network parameter to be optimized. \textcolor{black}{Along this direction, we learn the subsequent latent space $\hat{C}^{(t+2)}$,..., $\hat{C}^{(t+m)}$, from time $(t+2)$ to $(t+m)$ by the same pressure learning module:}
\begin{equation}
\hat{C}^{(t+2)}=Fpl(C^{(t-n+2)},...,C^{t}, \hat{C}^{{t+1}};\theta_f),
\end{equation}
\begin{equation}
\begin{split}
\hat{C}^{(t+m)}=Fpl(\hat{C}^{{(t-n+m)}},...,\hat{C}^{{(t+m-2)}}, \hat{C}^{{(t+m-1)}};\theta_f).
\end{split}
\end{equation}
The simplified process of the future pressure learning module is shown in Fig.~\ref{fig:lstm}, where we set $n=2$ and $m=4$ as an example. When $I^{t-1}$ and $I^t$ are fed into the $Fpl$ module, it can learn to predict $\hat{I}^{t+1}$. Similarly, when the input is $I^t$ and $\hat{I}^{t+1}$, the module can predict $\hat{I}^{t+2}$.

\subsection{Loss Function Design}
We have done a lot of experiments and optimization on the loss function design to ensure fast and accurate convergence of the MPMNet.
Traditionally, Mean Squared Error (MSE) and Mean Absolute Error (MAE) are used in standard data-driven methods by repeatedly calculating the difference between the output and ground truth values. A notable problem of MAE is that the gradient is considerably large, which may cause the minimum value to be missed at the end when optimizing the proposed model. For MSE, the gradient will gradually decrease as the loss value approaches its minimum. However, the MSE loss is susceptible to outliers, resulting in the reduced overall performance of the model. In these aspects, \textcolor{black}{the Huber loss function~\cite{huber:1992:REL}} can be beneficial because it reduces the gradient around the minimum and is also more robust to outliers than MSE. Therefore, it will fit our dataset better than MSE and MAE loss functions. To deal with the rate of pressure change, we also employ the Huber loss to compute the pressure gradient for the proposed network. Thus, the predicted loss $\mathcal{L}_{pre}$ is defined as:
\begin{equation}
\begin{split}
\mathcal{L}_{pre}=Huber(X^{f}, \hat{X}^{f})+Huber(\nabla{X^{f}}, \nabla{\hat{X}^{f}})\\
+Huber(X^{s}, \hat{X}^{s})+Huber(\nabla{X^{s}}, \nabla{\hat{X}^{s}})\\
+Huber(X^{i}, \hat{X}^{i})+Huber(\nabla{X^{i}}, \nabla{\hat{X}^{i}}),
\end{split}
\end{equation}

\begin{equation}
Huber\mathclap{(}X, \hat{X}\mathclap{)}=\left\{
\begin{alignedat}{2}
&\frac{1}{2}(X-\hat{X}\mathclap{)}^2, for|X(k)-\hat{X}(k)|\leq \theta_h,\\
&\theta_h|X-\hat{X}|-\frac{1}{2}\theta_h^2, otherwise
\end{alignedat}
\right.
\end{equation}
where $X$ denotes the ground truth data after the normalization process, and $\hat{X}$ is used to represent generated data from our prediction network. $X^f$, $X^s$, and $X^i$ denote fluid, solid pressure matrices, and the interface between fluid and solid bodies, separately. $\theta_h$ is the corresponding network parameters to be optimized.

We also train the autoencoder simultaneously, where $\mathcal{L}_{ae}$ is applied to the dataset pairs of autoencoder outputs $\tilde{X}$ and their corresponding ground truth $X$,
\begin{equation}
\begin{split}
\mathcal{L}_{ae}=Huber(X^{f}, \tilde{X}^{f})+Huber(\nabla{X^{f}}, \nabla{\tilde{X}^{f}})\\
+Huber(X^{s}, \tilde{X}^{s})+Huber(\nabla{X^{s}}, \nabla{\tilde{X}^{s}})\\
+Huber(X^{i}, \tilde{X}^{i})+Huber(\nabla{X^{i}}, \nabla{\tilde{X}^{i}}).
\end{split}
\end{equation}
The total loss is defined as:

\begin{equation}
\mathcal{L}_{total}=\mathcal{L}_{pre}+\mathcal{L}_{ae}.
\end{equation}
Note that the loss is computed over the entire trajectories in the training dataset. We jointly back-propagate through our model at every time step and tune the parameters to minimize the loss.

\subsection{Iterative Refinement}

Although end-to-end learning has the advantage of faster inference than numerical methods, it brings up the problem of error accumulation, which inevitably leads to the predicted pressure deviating from the actual value more and more as the prediction goes on, resulting in an unsmooth surface on the fluid and solid artifacts and even the wrong simulation for long-term prediction. Aware of these issues, our solution is to adopt a post-processing method with the help of a numerical solver to reduce the problems mentioned earlier. 

\begin{figure}[t]
\centering
\setlength{\abovecaptionskip}{0cm} 
\includegraphics[width=0.5\textwidth]{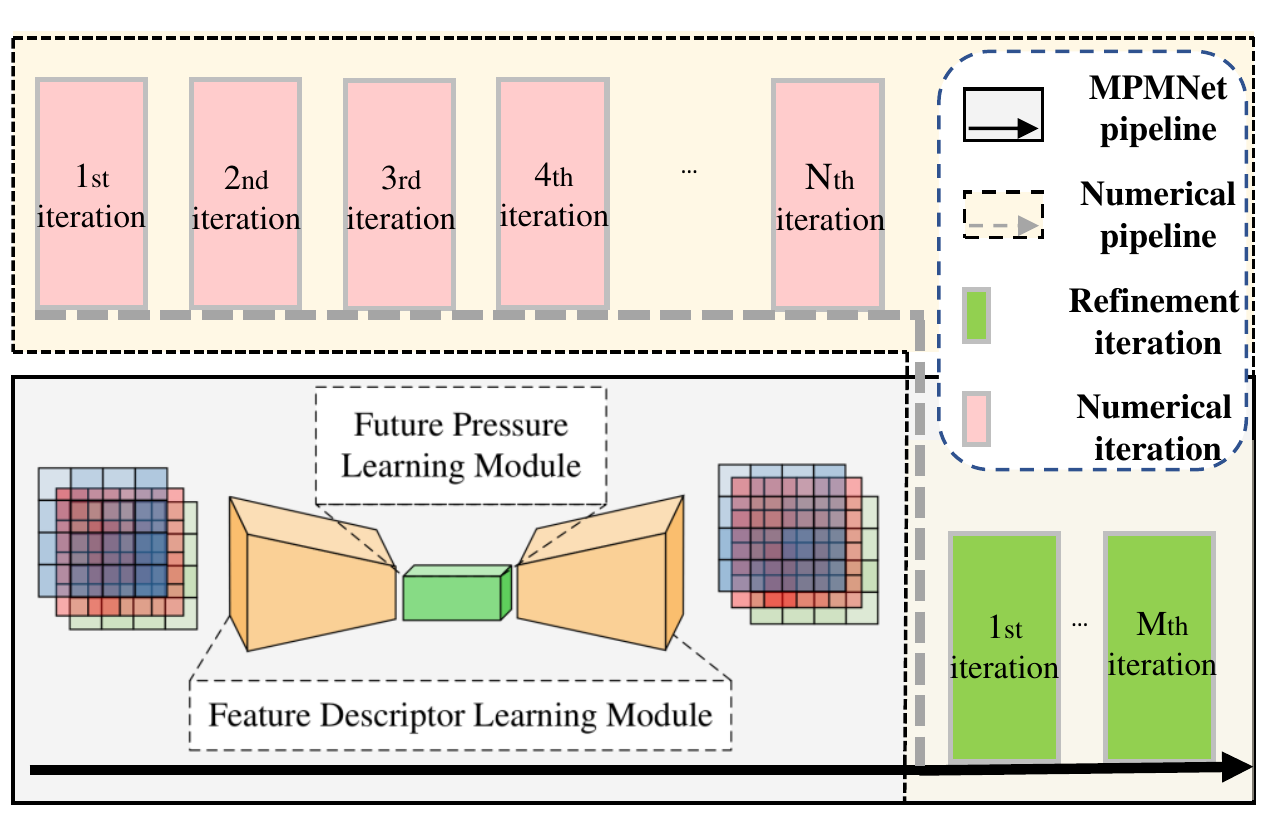}
\caption{Comparison of the traditional physics solver simulation with MPMNet. For the traditional solvers, we need $N+M$ iterations with $N$ numerical calculations and $M$ refinement iterations. For our MPMNet, we use the data-driven model to replace the beginning time-consuming $N$ iterations, and only a few $M$ refinements can achieve good performance.}
\label{fig:refinepipeline}
\end{figure}

In our method, the predicted pressure by networks will be mapped back to either the AMG solver or the G-S solver, then further modified by several iterations. The number of iterations is determined by the relative residual threshold (1e-3) of the iterative solvers. This combination process our prediction simulations can achieve higher stability and generalization, enabling long-term predictions of fluid-solid coupling and the stable prediction of the scene that does not appear in the training set. The predicted pressure $\hat{X}$ is inverse mapped through $\hat{\textbf{\emph{p}}}=Invmap(\hat{X},\textbf{\emph{r}})$ and then the field $\hat{\textbf{\emph{p}}}$ is computed iteratively as follows:
\begin{equation}
\hat{\textbf{\emph{p}}}=Refine(\hat{\textbf{\emph{p}}};\theta_r),
\label{equ:refine}
\end{equation}
where $\theta_r$ represents fluid-solid coupling simulation parameters in pressure iterations. Fig.~\ref{fig:refinepipeline} depicts the comparison of the traditional physics solver simulation with MPMNet.

Through these procedures, we deeply integrate the data-driven and physics-driven models by predicting pressures with a learning model and iteratively refining process with a numerical solver, thus significantly reducing iterative computations while maintaining physics-aware simulation.

\section{Experiments and Discussion}

\subsection{Network Training}
We use Pytorch$\footnote{https://www.pytorch.org}$ as our training framework. While acquiring training data, we set the relative convergence tolerance of the ground truth iteration method to 1e-3 to obtain an accurate dataset. In the data training step, the batch size for training is 8. Learning rate $l_r=0.0001$ is set as our initial value. Additionally, we use Adam\cite{kingma:2014:ADM} as our training optimizer to update the learning rate, and the max training iterations are 100,000. $n=4$ is set in the future pressure learning module.

We train our neural network model on a machine with a 3.80GHz Intel
i7-9800 16-core processor. It has 16 GB of memory and an NVIDIA RTX 2080Ti GPU card with 3584 CUDA cores and 11GB memory. The operating system is Ubuntu 18.04 OS. 

We got our trained network parameters after 5 hours of training (not including data generation time).
Our fluid-solid coupling experiment results are rendered with Redshift in \textit{Houdini}. It is important to note that we are not addressing the rendering issue in this article but rather focusing on simulation.

\subsection{Interacting Complexity}
For fair and clear comparisons, we introduce a new indicated parameter called "interacting complexity" to reflect the coupling complexity level of dynamic fluid-solid interaction. The interacting complexity is related to the characteristics of a dynamic fluid-solid scene, such as the number of dynamic solids ($\beta$), particles per cell ($\gamma$), the percentage of active grids (which involve dynamic particles) in the whole scene, the average velocity, etc. As a result, we can characterize the interacting complexity using a unified equation:
\begin{equation}\label{equ:iten}
\zeta
=\lg(\beta \times \gamma \times \frac{\sum^{}  \textbf{\emph{v}}^f+\sum^{} \textbf{\emph{v}}^s+\sum^{} \textbf{\emph{v}}^i}{d\times h\times w}),
\end{equation}
where $\textbf{\emph{v}}^f$, $\textbf{\emph{v}}^s$, $\textbf{\emph{v}}^i$ denote the velocity of the fluid, solid, and solid-fluid interface of grid cells, separately. $d\times h\times w$ denotes the resolution of the entire scene. And the $\lg$ function for normalizing the results. The higher value $\zeta$ means more violent interaction happens in the fluid-solid scene, so more iterations and larger calculations are needed from numerical solvers.

\subsection{Evaluation Methods}
To verify the accuracy of our MPMNet, we supply two quantitative methods of comparison to illustrate the reliability of the prediction results of our method.

Specifically, we employ the mean Peak Signal-to-Noise Ratio (PSNR) method based on pressure properties to conduct the numerical evaluation. We compute the PSNR of fluid and slip boundary pressure $e_{f}$, solid pressure $e_{s}$, and solid-fluid interface pressure $e_{i}$, respectively.
\begin{equation}
e_{f}=10\times\lg(\frac{\textbf{\emph{p}}^f_{max}}{\frac{1}{n_{1}}\sum_{k=0}^{n_{1}}||{\textbf{\emph{p}}^f}(k)-\hat{\textbf{\emph{p}}}^f(k)||^2}),
\end{equation}
\begin{equation}
e_{s}=10\times\lg(\frac{\textbf{\emph{p}}^s_{max}}{\frac{1}{n_{2}}\sum_{k=0}^{n_{2}}||{\textbf{\emph{p}}^s}(k)-\hat{\textbf{\emph{p}}}^s(k)||^2}),
\end{equation}
\begin{equation}
e_{i}=10\times\lg(\frac{\textbf{\emph{p}}^i_{max}}{\frac{1}{n_{3}}\sum_{k=0}^{n_{3}}||{\textbf{\emph{h}}}(k)-\hat{\textbf{\emph{h}}}(k)||^2}),\ \ \ 
\end{equation}
where $n_{1}$, $n_{2}$, $n_{3}$ denote the length of fields of fluid and slip boundary pressure, solid pressure, and solid-fluid interface pressure, respectively. The higher evaluation values mean better accuracy in our predictions. \textcolor{black}{Because the pressure does not change too much in the beginning of most interaction scenes, the measurements start after 50 steps of simulation and are averaged for ten scenes from the test setup.}

\textcolor{black}{Another critical issue is the max norm for velocity divergence. Using the max norm provides a measure of the largest deviation from the expected value and can be more informative in identifying regions with high divergence. So we compute the max norm for divergence of the fluid velocity at each step to measure the predicted accuracy using the following equation:
\begin{equation}
div=\max(|\nabla \cdot \hat{\textbf{\emph{v}}}^f|),
\end{equation}
where $\hat{\textbf{\emph{v}}}^f$ denotes the fluid velocity computed from the predicted pressure.} In the following, \textcolor{black}{ablation tests will be conducted} on different aspects of our method to evaluate their respective influence on the final results.

\subsection{Quantitative Evaluations}
\textcolor{black}{In this section, we compare our MPMNet with the baseline of previous works, and conduct ablation studies on different settings of our proposed architecture to analyze their respective influences on the output. The examples presented in the paper are newly generated and not included in the training set.} We compute the mean PSNR for our comparisons. We refer the readers to the supplemental video for the corresponding animations.

\begin{figure}[t]
\centering
\setlength{\abovecaptionskip}{0cm} 
\includegraphics[width=0.5\textwidth]{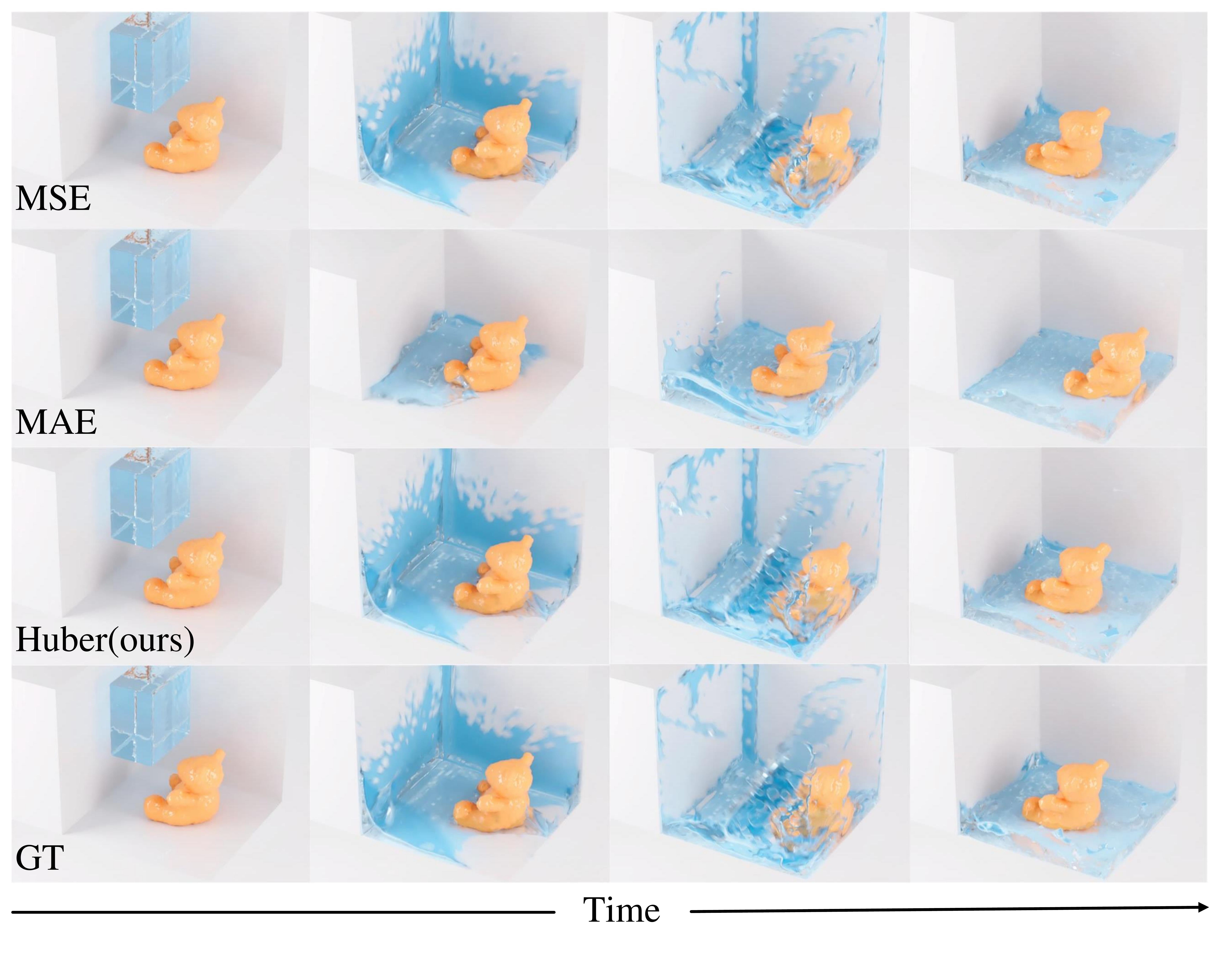}
\caption{Comparisons with MSE loss functions and MAE loss functions. The simulation results are generated by MSE loss functions, MAE loss functions, our method and the traditional physical method from top to bottom.}
\label{fig:huber}
\end{figure}

\textbf{\textit{Comparisons with different loss functions.}} The temporal awareness of our loss design is evaluated because of the significant impact on our temporal prediction network's performance. As shown in Fig.~\ref{fig:huber}, we have made comparisons of our loss function to the MSE and MAE loss functions. We have trained our model using these loss functions for 100,000 iterations. It shows that our proposed MPMNet model and MSE method can predict the motion and interaction of fluid and solids well, while the solid motion of the MAE method deviates from the real motion trajectory to varying degrees compared to ground truth. In Table~\ref{table:huber}, we can find that our MPMNet has a higher PSNR value of 33.31 for $e_f$, 28.42 for $e_s$, and 23.19 for $e_i$ when compared to other loss functions. The reason is that the huber loss reduces the gradient around the minimum and is more robust to outliers than MSE. Therefore, it will suit our dataset better than MSE or MAE loss functions. In contrast, our method employs loss synergies for pressure fields and accurately predicts future time steps.

\begin{table}[h]
\centering
\setlength{\abovecaptionskip}{0cm} 
\caption{Evaluations for loss functions}
\label{table:huber}
\begin{tabular}{@{}ccccc@{}}
\toprule
Loss Function & PSNR($e_f$) & PSNR($e_s$) & PSNR($e_i$) &  \\ \midrule
MSE            &  33.25           &  27.75          &   22.71          &  \\
MAE            &  29.31           &  25.29          &   19.72          &  \\
Huber(ours)    &  \textbf{33.31}  &  \textbf{28.42} &   \textbf{23.19} &  \\ \bottomrule
\end{tabular}
\end{table}

\textbf{\textit{Comparisons with the prediction window size.}} The prediction window size $n$ of consecutive time steps that are taken as input by the future pressure learning module has a major impact on the resulting weight count of the network as well as the efficiency of the temporal predictions and thereby their difficulty. In the following, we have compared the size $n$ of 3, 4 and 5. The results in terms of PSNR values are displayed in Table~\ref{table:input}. It is clear that the result benefits from $n = 4$. For this reason, we use a prediction window 4 for all subsequent comparisons.

\begin{table}[]
\centering
\setlength{\abovecaptionskip}{0cm} 
\caption{Evaluations for the prediction window size $n$}
\label{table:input}
\begin{tabular}{@{}ccccc@{}}
\toprule
$n$       & PSNR($e_f$) & PSNR($e_s$) & PSNR($e_i$) &  \\ \midrule
3         &  29.03           & 24.62          & 17.41          &  \\
4         &  \textbf{29.35}  & \textbf{24.91} & \textbf{17.62}          &  \\
5         &  29.12           & 24.18          & 17.49          &  \\ \bottomrule
\end{tabular}
\end{table}

\begin{figure}[htb]
\centering
\includegraphics[width=0.5\textwidth]{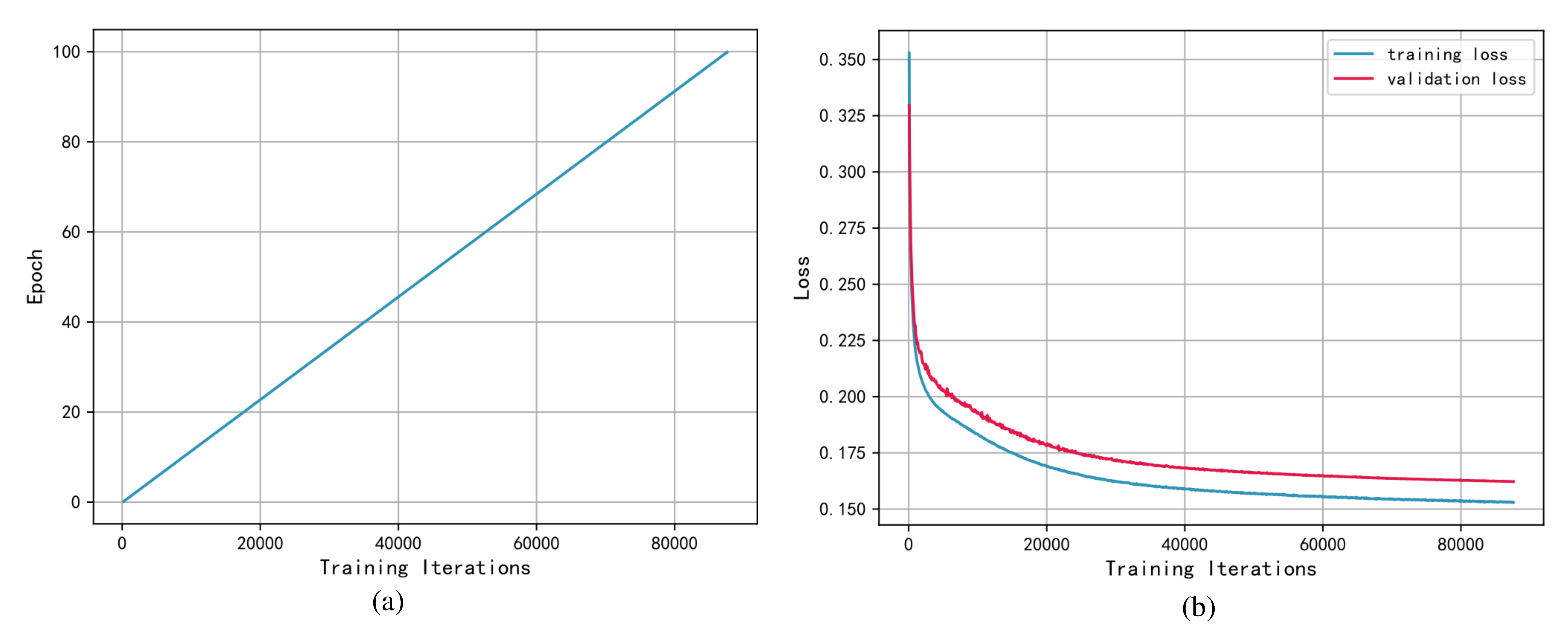}
\caption{\textcolor{black}{Convergence plots for our training process. (a) is for training epochs; (b) is for our method.}}
\label{fig:cplot-our}
\end{figure}

\begin{figure}[htb]
\centering
\includegraphics[width=0.5\textwidth]{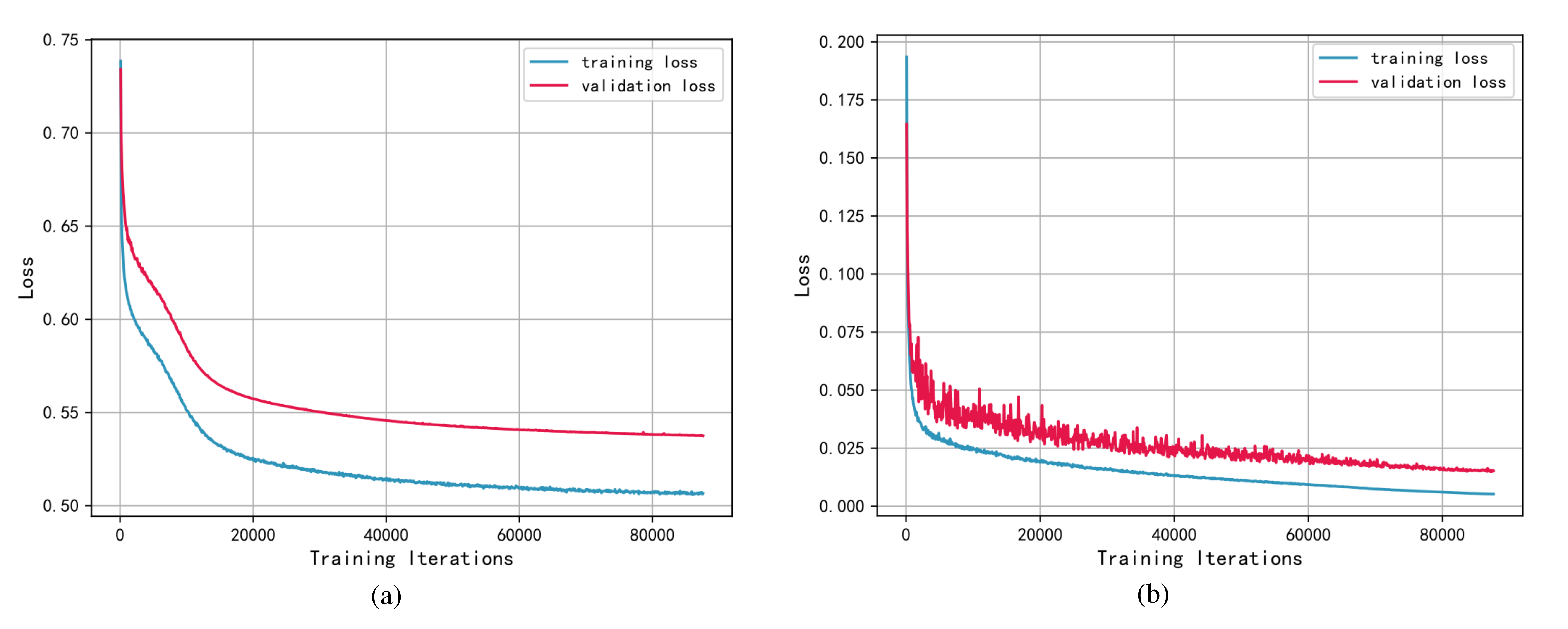}
\caption{\textcolor{black}{Convergence plots for SOTA methods. (a) is for Latent Space Physics method~\cite{Wiewel:2019:LSP}; (b) is for Deep Fluids method~\cite{kim:2019:DFG}.}}
\label{fig:cplot-DL}
\end{figure}

\textbf{\textit{Comparisons with SOTA methods.}} \textcolor{black}{We have conducted two comparison experiments with the Latent Space Physics method (Wiewel et al.~\cite{Wiewel:2019:LSP}) and Deep Fluids method (Kim et al.~\cite{kim:2019:DFG}). To ensure fair comparisons, we adopt the same network architecture as our method but with different input contents.
For the comparison with the Latent Space Physics method, we feed three channels with the same pressure field of the entire domains without distinction.
For the comparison with the Deep Fluids method, we feed three channels with the velocity field. As shown in Fig.~\ref{fig:cplot-our} (a), we have trained 100 epochs for our method, Latent Space Physics and Deep Fluids approach, separately. Fig.~\ref{fig:cplot-our} (b) shows the convergence plot of our training process. Fig.~\ref{fig:cplot-DL} shows the training convergence plots of Latent Space Physics and Deep Fluids approach.} As shown in Table~\ref{table:lsp}, our method achieves a higher PSNR, confirming that we have a clear preference for the interaction learning of fluids and solids.

\begin{table}[h]
\centering
\setlength{\abovecaptionskip}{0cm} 
\caption{Evaluations of the Latent Space Physics method, Deep Fluids method, and our approach}
\label{table:lsp}
\setlength{\abovecaptionskip}{0pt}
\begin{tabular}{@{}ccccc@{}}
\toprule
Approach             & PSNR($e_f$)      & PSNR($e_s$) & PSNR($e_i$) &  \\ \midrule
Latent Space Physics~\cite{Wiewel:2019:LSP}&  26.84           & 19.68          & 15.39          &  \\
Deep Fluids~\cite{kim:2019:DFG}          &25.42                  &19.25                &15.31         &  \\
ours                 &  \textbf{29.73}  & \textbf{25.73} & \textbf{18.50} &  \\ \bottomrule
\end{tabular}
\end{table}

\textbf{\textit{Comparisons with different refinements.}} We utilize AMG and G-S method as the post-processing step to refine the pressure further. Table~\ref{table:refine} shows that both methods get similar levels of results. But the two iterative methods consume different computation times. For simple scenarios, we choose the G-S method to obtain less refinement time, and correspondingly, AMG iterative refinement is more suitable for complex scenarios.


\begin{table}[]
\centering
\setlength{\abovecaptionskip}{0cm} 
\caption{Evaluations for the different refinements}
\label{table:refine}
\begin{tabular}{@{}ccccc@{}}
\toprule
Refinement  & PSNR($e_f$) & PSNR($e_s$) & PSNR($e_i$) &  \\ \midrule
G-S         &  30.64      & 28.35       & 25.63          &  \\
AMG         &  30.93      & 28.21       & 25.56 &  \\ \bottomrule
\end{tabular}
\end{table}

\subsection{Velocity Divergence Evaluations}
\begin{figure}[htbp]
\centering
\setlength{\abovecaptionskip}{0cm} 
\includegraphics[width=0.5\textwidth]{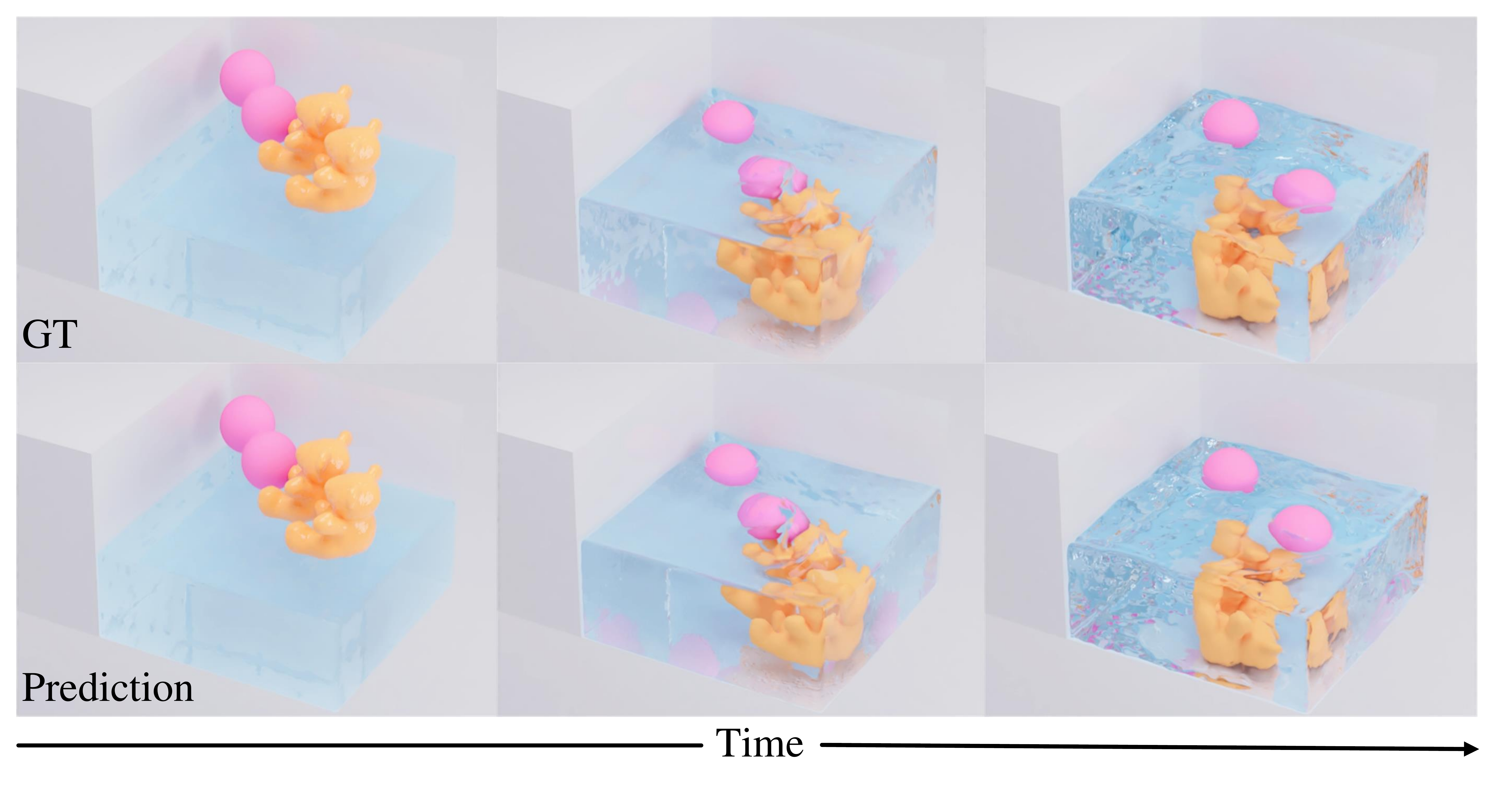}
\caption{Solids of different densities interact with water.}
\label{fig:density}
\end{figure}

\begin{figure}[htb]
\centering
\includegraphics[width=0.5\textwidth]{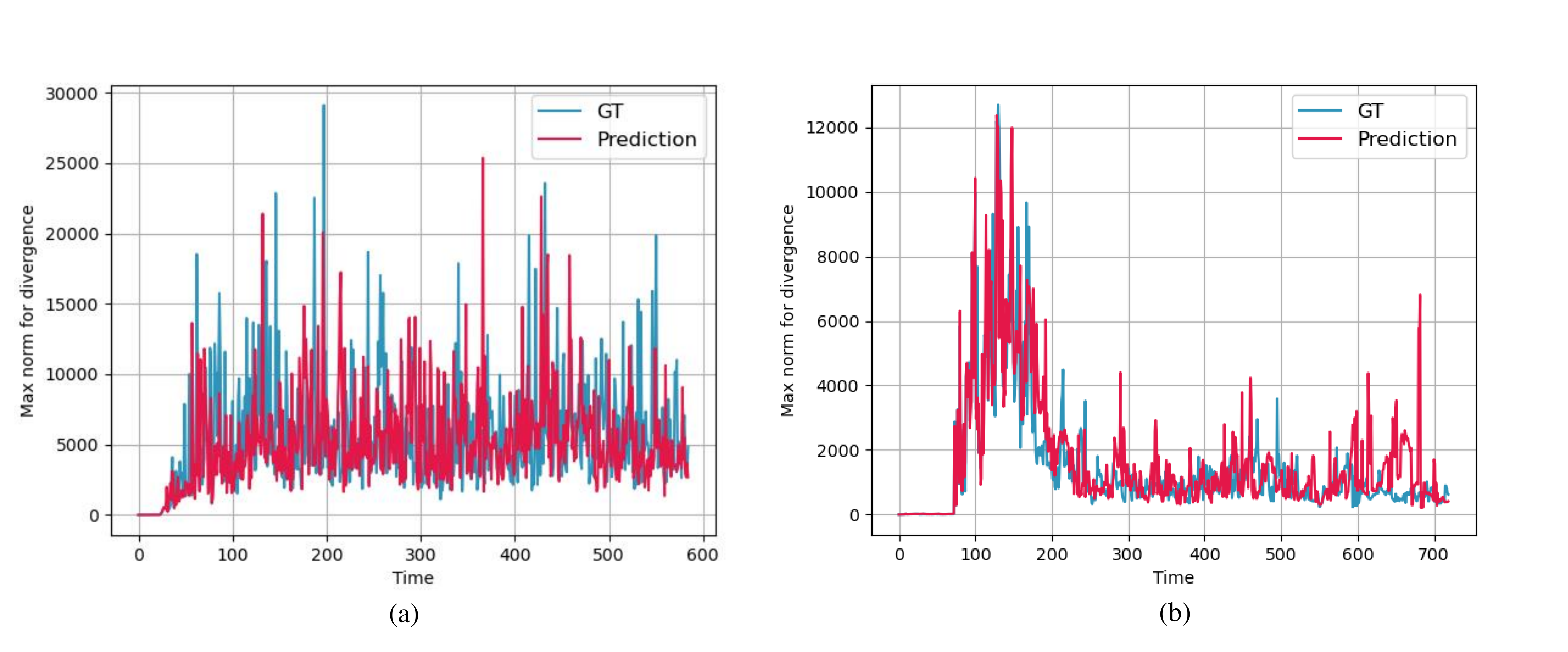}
\caption{\textcolor{black}{Comparisons of the divergence between ground truth
velocity and predicted velocity. (a) shows the velocity divergence of Fig.~\ref{fig:density}; (b) shows the velocity divergence of Fig.~\ref{fig:generation}.}}
\label{fig:refinemax}
\end{figure}

Velocity divergence-free can constraint the incompressibility of fluid. Fig.~\ref{fig:density} shows solids of different densities can interact differently with water, illustrating the ability of our method to learn the physical properties of solids. \textcolor{black}{In Fig.~\ref{fig:refinemax} (a), we have compared the prediction results with ground truth for the max norm for velocity divergence. The line graph shows that our method achieves good fluid incompressibility as the ground truth, which concludes that our MPMNet has good adaptability and can be conducted with different numerical solvers.}

\begin{figure*}[htbp]
\centering
\setlength{\abovecaptionskip}{0cm} 
\includegraphics[width=1.0\textwidth]{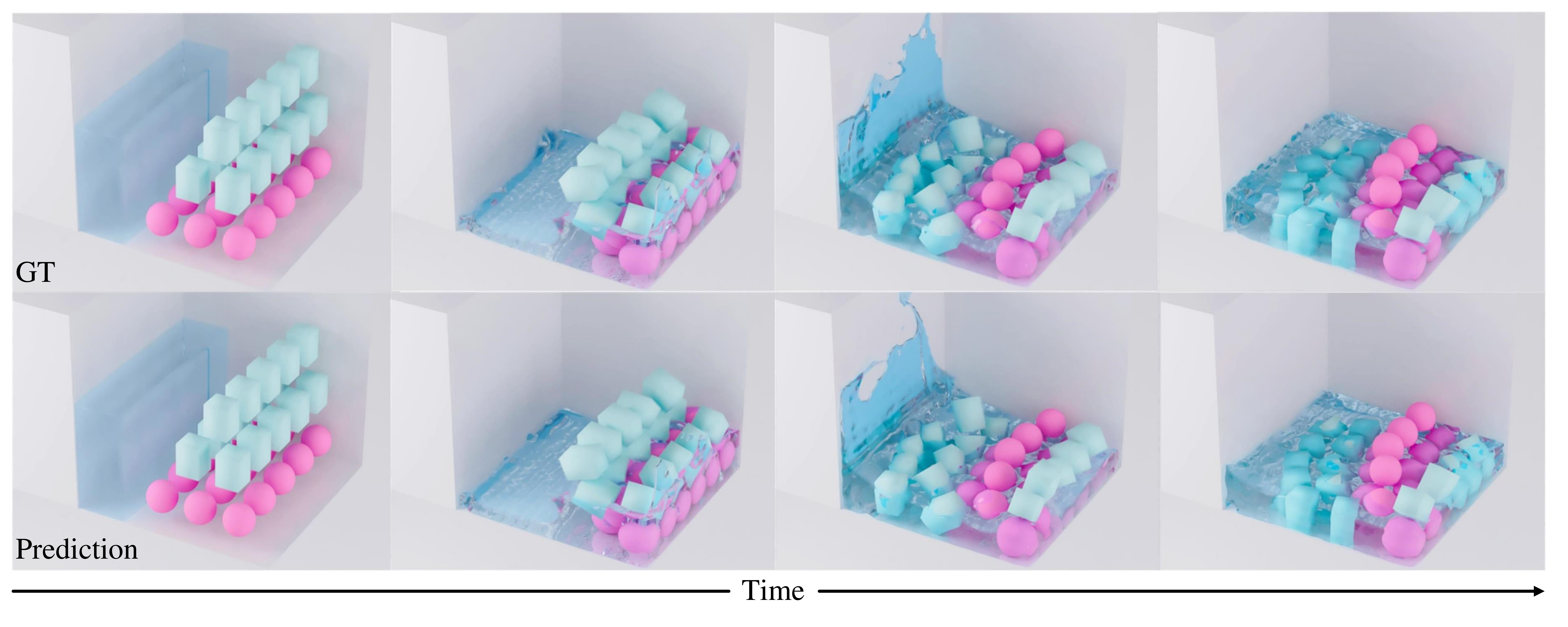}
\caption{Predicted results of water block flushing 30 solids using our method.}
\label{fig:generation}
\end{figure*}

We have further attempted to predict more complex scenarios not contained in the training scenes. Fig.~\ref{fig:generation} displays the predicted results of water block flushing 30 solids using our MPMNet. It shows that our method can better approximate the real situation and maintain long-term stability. \textcolor{black}{As can be seen from Fig.~\ref{fig:refinemax} (b), even after long-term prediction, our divergence constraint is still very close to the ground truth, which can ensure good accuracy and incompressibility of fluid. At the same time, we have analyzed all the divergence in each frame and found that 99.99\% of them were within 3. Only a few have large divergences. These values exist at fluid-solid interface and fluid free surfaces.}

\subsection{New Scenes Generalization}
\begin{figure*}[htbp]
\centering
\setlength{\abovecaptionskip}{0cm} 
\includegraphics[width=1.0\textwidth]{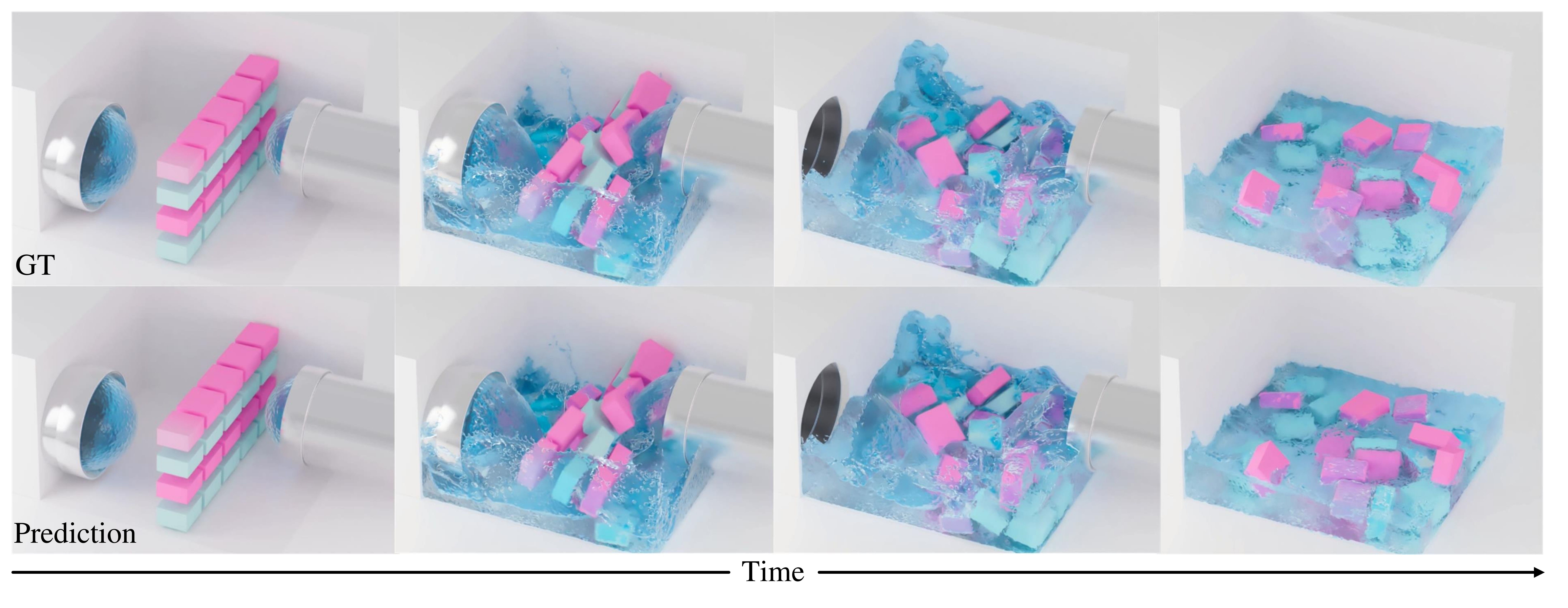}
\caption{Prediction results at the resolution of $64^3$. Emitters impact solids.}
\label{fig:emitter}
\end{figure*}

\begin{figure*}[htbp]
\centering
\setlength{\abovecaptionskip}{-0.2cm} 
\includegraphics[width=1.0\textwidth]{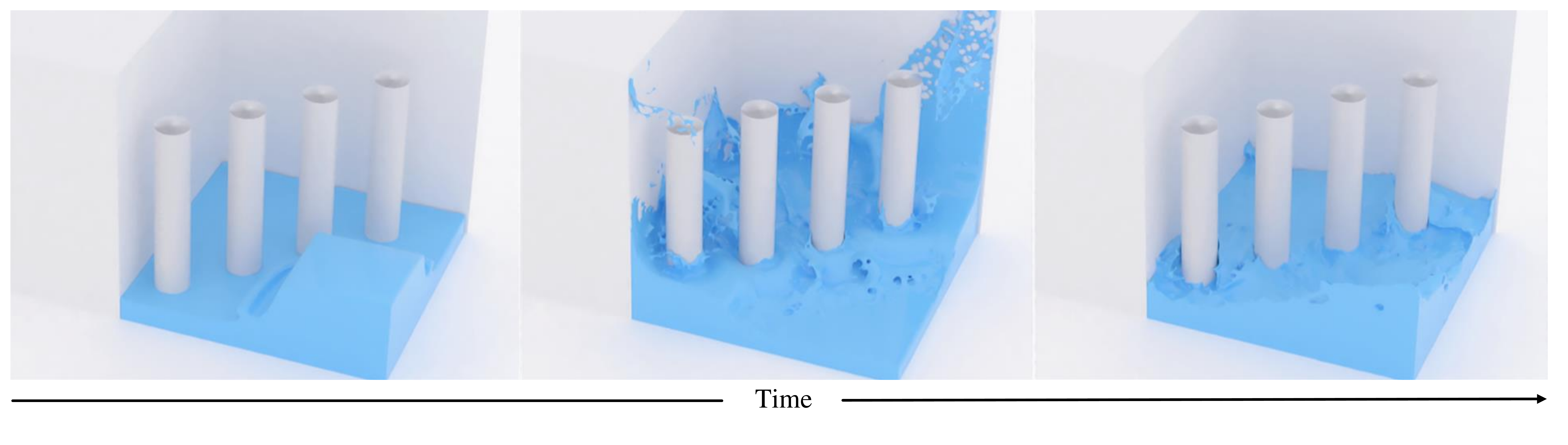}
\caption{Prediction results at the resolution of $128^3$. Water block impacts four static cylinders.}
\label{fig:cylinder}
\end{figure*}

\begin{figure*}[htbp]
\centering
\setlength{\abovecaptionskip}{-0.2cm} 
\includegraphics[width=1.0\textwidth]{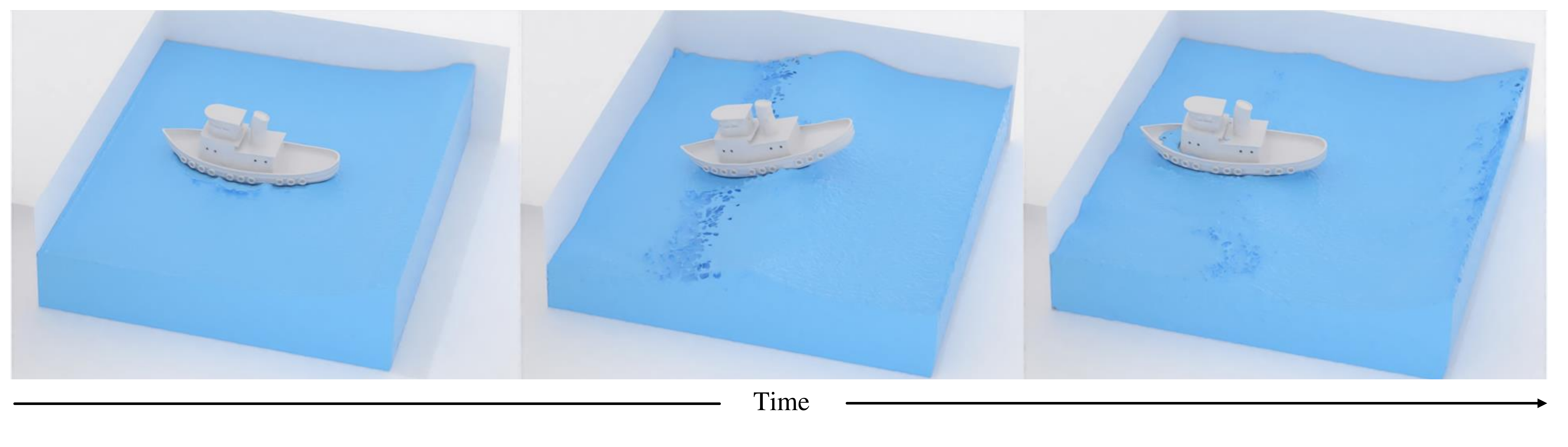}
\caption{\textcolor{black}{Prediction results at the resolution of $128^3$. A Boat floating in the waves.}}
\label{fig:boat}
\end{figure*}

We have also evaluated our framework on various resolutions to prove that our method produces visually-plausible results at a fast speed. Note that since our network is fully-convolutional, the size of the domain can be modified at inference time.
As shown in Fig.~\ref{fig:emitter}, we have made our prediction at the resolution of $64^3$. Fig.~\ref{fig:cylinder} and Fig.~\ref{fig:boat} show the prediction results of our method at $128^3$ resolution.
Despite the fact that the generalization scenes are not part of the training data, our network performs quite well in all testing tasks with rich details and violent interactions.

\subsection{Performance Analysis}
Table~\ref{speedTable} summarizes the statistics of all presented examples, including scene settings, traditional computation times by AMG iteration and G-S iteration, inference time, and refinement time. The time of our method is the sum of inference time and refinement time. The AMG iterations, G-S iterations, and our network inference are all done by CUDA on the same computer system. And Table~\ref{table:iten} summarizes the interacting complexity of all presented scenes. Exceptionally, for one-way static interactions,
such as the scene "Cylinders Dam" (Fig.~\ref{fig:cylinder}), there is no dynamic solids, we set $\beta$ to $0.1$ as a minimum value for reasonableness. However, because our MPMNet is not affected by interacting complexity, we can drastically reduce numerical computation time. In terms of wall clock time, our proposed approach generates desired pressure fields up to 28$\times$ faster than simulating the data with the pure physical method. As the "interacting complexity" increases, the inference time of our network does not change at the same resolution, so our method can gain a more obvious acceleration advantage.

\begin{table*}[htbp]
\renewcommand{\arraystretch}{1.3}
\setlength{\abovecaptionskip}{0cm} 
\caption{Statistics and time costs for our MPMNet and iteration methods. Note that the computation, AMG iterations, G-S iterations, and network inference are all done by CUDA on the GPU. Unless otherwise stated, the AMG and G-S residual threshold is set to 1e-3}
\label{speedTable}
\centering
\begin{tabular*}{\linewidth}{@{}ccccccccccc@{}}
\cmidrule(r){1-9}
Scene& \makecell[c]{Grid\\ Resolution}& \# Frames & \makecell[c]{Computation \\Time(AMG,ms)}& \makecell[c]{Computation \\Time(G-S,ms)}& \makecell[c]{Inference \\Time(ms)}& \makecell[c]{Refinement \\Time(ms)} & \makecell[c]{Interacting\\ Complexity}& \textbf{\makecell[c]{Speed Up\\($\times$)}} \\ \cmidrule(r){1-9}
Bear Bath (Fig.~\ref{fig:huber})           & $32^3$    & 585   & 53.73                         & 177.03   & 4.26      & 6.82(G-S)       &2.15    & \textbf{4.85}                \\
Solids Drop (Fig.~\ref{fig:density})         & $32^3$    & 585   & 82.03                         & 103.32   & 4.31     & 8.2(G-S)       & 2.79    & \textbf{6.56}                \\
Solids Dam (Fig.~\ref{fig:generation})    & $32^3$   & 720   & 901.25                        &1192.69  & 4.33      & 27.45(AMG)      &3.97 &\textbf{28.36}               \\
Emitter (Fig.~\ref{fig:emitter})            & $64^3$   & 675   & 558.79 
&984.26   & 15.13     & 53.66(AMG)      &3.06      & \textbf{8.12
} \\
Cylinders Dam (Fig.~\ref{fig:cylinder})       & $128^3$  & 900   & 1136.09 
&1583.63  & 88.32     & 482.50(AMG)    &  0.45   & \textbf{1.99
} \\
Boat (Fig.~\ref{fig:boat})                 & $128^3$  & \textcolor{black}{1620}  & \textcolor{black}{872.40}                        &\textcolor{black}{1003.15}   & \textcolor{black}{87.93}     & \textcolor{black}{297.62(AMG)}      &  1.24 & \textbf{\textcolor{black}{2.26}} \\\bottomrule
\end{tabular*}
\end{table*}

\begin{table}[htbp]
\centering
\renewcommand{\arraystretch}{1.3}
\setlength{\abovecaptionskip}{0cm} 
\scriptsize
\caption{Evaluations of the complexity of the fluid-solid interaction}
\label{table:iten}
\setlength{\tabcolsep}{0.9mm}
\begin{tabular}{@{}cccccc@{}}
\cmidrule(r){1-6}
Scene           &\makecell[c]{Dynamic \\Solids $\beta$} & \makecell[c]{Particles \\
per Cell $\gamma$} & \makecell[c]{ Active \\ Grids (\%)} & \makecell[c]{Average \\Velocity} & \makecell[c]{\textbf{ Interacting}\\  \textbf{Complexity $\zeta$}}  \\ \cmidrule(r){1-6}
Bear Bath  (Fig.~\ref{fig:huber})     &1    &$16^3$&0.11    &0.31  &\textbf{2.15} \\
Solids Drop (Fig.~\ref{fig:density})   &3    &$16^3$&0.48    &0.11  &\textbf{2.79}\\ 
Solids Dam (Fig.~\ref{fig:generation})   &30  &$16^3$ &0.28   &0.27  &\textbf{3.97}\\
Emitter (Fig.~\ref{fig:emitter})        &20   &$8^3$  &0.26   &0.43  &\textbf{3.06}\\
Cylinders Dam (Fig.~\ref{fig:cylinder})  &0.1  &$8^3$ &0.25   &0.22  &\textbf{0.45}\\
Boat (Fig.~\ref{fig:boat})          &1    &$8^3$  &0.21  &0.16  &\textbf{1.24
}\\
\bottomrule
\end{tabular}
\end{table}

\subsection{Discussion and Limitation}

The above-documented experimental series and corresponding data
analysis show that for fluid-solid scene reconstructions with
different physical properties, scene scales, and shapes, our method
can \textcolor{black}{consistent} achieve convincing visual effects and ensure consistent accuracy
w.r.t. the numerical methods. The core contribution of our method
is to reduce time expenses by lowering a large number of initial
numerical iterations while simultaneously designing data-driven and
physics-driven combined networks to maintain a sense of physical
reality in fluid-solid interactions. 

In practice, as the interacting complexity increases, the time consumption required by numerical
iteration increases exponentially. Since our MPMNet is not sensitive to the dynamic complexity, the time
expense for pressure prediction will not increase as much, so our
method's advantage is expected to become more prominent with more
complicated interactions. On the contrary, due to the same rationale, we can hardly demonstrate the superiority of such applications with simple interacting intensities or low requirements on numerical precision.

Meanwhile, although we have shown that our method can predict visually
similar results, there are still some limitations that call for
possible improvements in the future. First, though MPMNet has the ability to generate behaviors that do not exist in the training data, it can hardly produce ideal results when the input parameters significantly differ from the training data. This is a common problem with
data-driven tasks that are always constrained by raw data and could be
alleviated by either significantly expanding the scope of datasets or
altering network architecture with potentially different design
philosophies. In addition, our model only deals with fluid and solids'
interaction/motion and did not yet include the specific design for
deformable bodies or soluble solids. Some subtle deformations that do
not cause dramatic changes in pressure can be learned naturally with
our framework, but the large deformations cannot be handled yet.

\section{Conclusion and Future Work}

In this paper, for the first time, we propose a novel data-driven and
physics-driven hybrid fluid-solid prediction method that applies neural networks to MPM for liquid-solid interactions in computer graphics. Our immediate goal is to improve simulation
speed while maintaining good stability, generalization, and
versatility. Our cooperative model, the MPMNet, can simulate fluid-solid coupling
results by sensing different material behaviors and physical features
with corresponding data-driven components while still trying to be as
visually and numerically accurate as what the conventional
physics-based approaches can achieve. Compared with the traditional
approaches, our method greatly improves the solution efficiency of
dynamic fluid-solid interaction while using pressures of only a few
frames from the involved numerical solver. Meanwhile, we introduce an interacting complexity value to indicate the complexity of fluid-solid interactions. 
The comprehensive experiments verify that our approach has tremendous potential and advantages for greater acceleration as the complexity increases.

Looking beyond the current accomplishment, we plan to further explore
the network's potential to learn physical processes to enable our
method to handle fluid-solid interactions involving different
properties, such as viscous fluid, deformable solids, and multiphase
flow interactions. \textcolor{black}{Regarding other potential applications, we believe that the use of pressure matrix could also be explored for other topics of phenomena modeling, e.g., the use of viscosity matrix projections for non-Newtonian materials simulation. Moreover, the use of matrix projections is not exclusive to the MPM method, this could also be integrated with other MPM-related methods such like APIC/FLIP/PIC. We hope that our work can inspire further research in these directions and contribute to the new development of more powerful and versatile data-driven approaches towards physics-driven simulation.}

\ifCLASSOPTIONcaptionsoff
  \newpage
\fi



%

\bibliographystyle{IEEEtran}
\bibliography{IEEEabrv,template}


\end{document}